\documentclass[nonacm,sigconf,review=false,anonymous=false]{acmart}

\usepackage{enumitem}
\usepackage{subcaption}
\usepackage{multirow}
\usepackage{amsmath}
\usepackage{booktabs}

\renewcommand\footnotetextcopyrightpermission[1]{} 
\AtBeginDocument{%
  }

\begin{document}

\title{LiquidGEMM: Hardware-Efficient W4A8 GEMM Kernel for High-Performance LLM Serving}

\author{Huanqi Hu}
\authornote{Both authors contributed equally to this research.}
\affiliation{%
  \institution{Shanghai Jiao Tong University}
  \city{Shanghai}
  \country{China}
}
\email{huhuanqi@sjtu.edu.cn}

\author{Bowen Xiao}
\authornotemark[1]
\affiliation{%
  \institution{ByteDance Seed}
  \city{Shanghai}
  \country{China}
}
\email{xiaobowen.96@bytedance.com}

\author{Shixuan Sun}
\authornote{Shixuan Sun is the corresponding author.}
\affiliation{%
  \institution{Shanghai Jiao Tong University}
  \city{Shanghai}
  \country{China}
}
\email{sunshixuan@sjtu.edu.cn}

\author{Jianian Yin}
\affiliation{%
  \institution{ByteDance Seed}
  \city{Shanghai}
  \country{China}
}
\email{yinjianian.0@bytedance.com}

\author{Zhexi Zhang}
\affiliation{%
  \institution{ByteDance Seed}
  \city{Shanghai}
  \country{China}
}
\email{zhangzhexi@bytedance.com}

\author{Xiang Luo}
\affiliation{%
  \institution{ByteDance Seed}
  \city{Shanghai}
  \country{China}
}
\email{luoxiang.17@bytedance.com}

\author{Chengquan Jiang}
\affiliation{%
  \institution{ByteDance Seed}
  \city{Seattle}
  \country{United States of America}
}
\email{jiangchengquan@bytedance.com}

\author{Weiqi Xu}
\affiliation{%
  \institution{ByteDance Seed}
  \city{Beijing}
  \country{China}
}
\email{xuweiqi117@gmail.com}

\author{Xiaoying Jia}
\affiliation{%
  \institution{ByteDance Seed}
  \city{Beijing}
  \country{China}
}
\email{jiaxiaoying@bytedance.com}

\author{Xin Liu}
\affiliation{%
  \institution{ByteDance Seed}
  \city{Seattle}
  \country{United States of America}
}
\email{liuxin.ai@bytedance.com}

\author{Minyi Guo}
\affiliation{%
  \institution{Shanghai Jiao Tong University}
  \city{Shanghai}
  \country{China}
}
\email{guo-my@cs.sjtu.edu.cn}

\renewcommand{\shortauthors}{Huanqi Hu et al.}

\begin{abstract}
Quantization is a critical technique for accelerating LLM inference by reducing memory footprint and improving computational efficiency. Among various schemes, 4-bit weight and 8-bit activation quantization (W4A8) offers a strong balance between accuracy and performance. However, existing W4A8 GEMM kernels fall short in practice due to inefficient dequantization on CUDA Cores, which cannot keep pace with the high throughput of Tensor Cores. In this paper, we present LiquidGEMM, a hardware-efficient W4A8 GEMM kernel for efficient LLM serving. LiquidGEMM designs two key techniques: LiquidQuant, a hardware-efficient quantization method that enables fast, overflow-safe dequantization using just two arithmetic instructions per four elements; and an implicit fine-grained pipeline that fully overlaps weight loading, dequantization, and MMA across warp groups without software synchronization or redundant memory traffic. Experimental results show that LiquidGEMM achieves up to 2.90x speedup over state-of-the-art W4A8 kernels and up to 4.94x end-to-end system-level speedup. Compared to various quantized GEMM kernels in NVIDIA TensorRT-LLM, LiquidGEMM delivers 1.12-1.63x performance gains, and achieves up to 1.63x system-level speedup.

\end{abstract}

\maketitle

\section{Introduction} \label{sec:introduction}

LLMs have transformed a wide range of applications, from natural language understanding to content generation, significantly advancing the capabilities of AI. However, their massive model size and computational intensity pose serious challenges for efficient deployment in production environments. To mitigate these issues, \emph{integer quantization}~\cite{dettmers2022gpt3,xiao2023smoothquant,frantar2022gptq,lin2024awq,wang2023bitnet, xu2024onebit} has emerged as a key technique. By converting full-precision floating-point values (FP32 or FP16) into low-precision integer formats (e.g., INT4), it reduces model size, lowers memory bandwidth requirements, and accelerates inference on hardware optimized for low-precision arithmetic.

\begin{figure}[t]\small
    \setlength{\abovecaptionskip}{3pt}
    \setlength{\belowcaptionskip}{0pt}
    \includegraphics[scale=0.65]{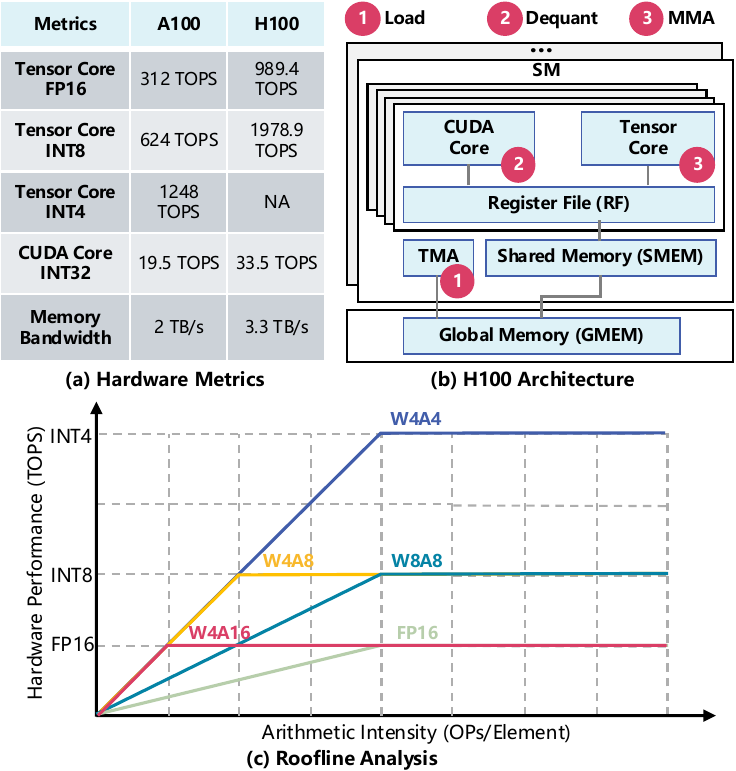}
    \centering
    \caption{Key performance metrics of NVIDIA A100 and H100 GPUs, with the roofline for GEMM layers in LLM serving.}
    \label{fig:roofline}
\end{figure}

Among various quantization configurations, recent studies~\cite{lin2024qserve,zhang2024qqq} highlight 4-bit weight and 8-bit activation quantization (\textbf{W4A8}) as a compelling trade-off between accuracy, efficiency, and memory usage. As illustrated in the roofline analysis (Figure~\ref{fig:roofline}), W4A8 outperforms \textbf{W4A16} by exploiting the high throughput of low-bit Tensor Core operations, delivering better performance in compute-bound scenarios such as large-batch inference. Compared to \textbf{W8A8}, W4A8 not only reduces memory footprint but also lowers memory bandwidth requirements, making it particularly advantageous in memory-bound settings like small-batch inference. Additionally, W4A8 improves arithmetic intensity, thereby reducing the batch size needed to saturate GPU compute resources. While more aggressive configurations such as \textbf{W4A4} offer similar model compression, they often incur substantial accuracy degradation due to heavily quantized activations~\cite{lin2024qserve,zhang2024qqq}. In contrast, W4A8 maintains higher accuracy while preserving most of the efficiency benefits. Due to its advantages, W4A8 quantization is a promising solution for high-performance LLM serving in production environments.

\setlength{\textfloatsep}{0pt}

GEMM (General Matrix Multiplication) operations are the core computational building blocks in LLM serving and critically influence inference efficiency. However, our experiments show that the state-of-the-art W4A8 GEMM implementation~\cite{lin2024qserve} fails to meet expectations: it does not outperform higher-precision methods like W8A8 in memory-bound scenarios and is significantly slower than W8A8 and even FP16 in compute-bound regimes, e.g., on LLaMA2-7B with a batch size of 256, the existing W4A8 GEMM~\cite{lin2024qserve} is 2x slower than W8A8. This contradicts recent roofline analyses~\cite{lin2024qserve,zhang2024qqq}, which suggest that W4A8 should outperform W8A8 in memory-bound cases and achieve comparable performance in compute-bound settings.

To analyze the problem, we profile the W4A8 dequantization method and develop a cost model that captures key performance factors in pipelined GEMM execution (see Section~\ref{sec:motivation} for details). Our analysis shows that the core issue lies in the hardware-unaware dequantization step preceding MMA, which incurs significant overhead due to the large performance gap between CUDA Cores and Tensor Cores. As illustrated in Figure~\ref{fig:roofline}b, after loading weights from GMEM to RF, W4A8 GEMM must first dequantize 4-bit weights to 8-bit on CUDA Cores before executing MMA (Matrix Multiply-Accumulate) on Tensor Cores. The QoQ algorithm~\cite{lin2024qserve}, which dequantizes multiple elements within a 32-bit register, suffers from potential overflow and requires dozens of instructions to resolve it. This imposes substantial compute pressure on the limited-capacity CUDA Cores, which cannot keep up with the high-throughput Tensor Cores (Figure~\ref{fig:roofline}a). In contrast, W8A8 GEMM avoids this dequantization step before MMA and can fully exploit Tensor Core performance. Consequently, although W4A8 is theoretically promising, the dequantization step becomes a performance bottleneck that limits its practical efficiency. This gap between roofline potential and actual performance underscores a fundamental limitation of current W4A8 methods.

To tackle the fundamental bottleneck in W4A8 LLM serving, we propose \textbf{LiquidGEMM}, a hardware-efficient W4A8 GEMM kernel for high-performance LLM serving. LiquidGEMM enables pipeline-parallel execution across heterogeneous GPU hardware units including TMA, CUDA Cores, and Tensor Cores to overlap dequantization with weight loading and MMA, thereby hiding its overhead and maximizing hardware utilization. Achieving this requires addressing two key challenges. First, the quantization algorithm must be optimized to reduce the computational burden of dequantization on CUDA Cores, which have limited compute throughput compared to Tensor Cores, so that it can be effectively overlapped with other stages. Second, the execution pipeline must be co-designed to coordinate data movement and computation efficiently, feeding the beast of Tensor Cores, the primary compute engine for GEMM on modern GPUs.

To address these challenges, we first propose \emph{LiquidQuant} (LQQ), a hardware-efficient W4A8 quantization scheme designed for native support by GPU instructions. Unlike prior methods that directly quantize INT8 to UINT4, leading to overflow issues during dequantization, LQQ applies a rotation-based transformation that shifts INT8 values into the UINT8 range before quantizing to UINT4. Paired with this rotation, we design a sweet dequantization strategy that leverages the properties of \emph{two’s complement representation} to recover the original INT8 values entirely within the UINT8 domain, without overflow. This dequantization is highly hardware-efficient, requiring only two 32-bit hardware instructions—\texttt{IMAD} and \texttt{XOR}—to process four elements, significantly reducing the computational load on CUDA Cores.

Next, we design the \emph{implicit fine-grained pipeline} (ImFP) execution mechanism for LiquidGEMM. On NVIDIA Hopper GPUs, a straightforward extension to existing warp-specialized GEMM pipelines is to assign an additional warp group (WG) for dequantization, aiming to overlap weight loading, dequantization, and MMA. However, this method incurs significant overhead from round-trip data movement between RF and SMEM for WG communication and costly inter-warp synchronization, resulting in pipeline bubbles and reduced efficiency. To address the problem, our ImFP adopts a single-producer, multiple-consumer execution model. A dedicated \emph{Load WG} transfers weights from GMEM to SMEM, and the GEMM workload is partitioned into fine-grained tasks that are dynamically consumed by multiple \emph{Compute WGs} in a preemptive manner. Each Compute WG immediately performs MMA on the weights it has dequantized, eliminating the round-trip data movement between SMEM and RF. Overlapping of dequantization and MMA is achieved across concurrently executing Compute WGs. Notably, task scheduling is managed by hardware, thereby avoiding the overhead of software synchronization. Centered on this pipeline design, we further optimize data layout and dequantization. LiquidGEMM is currently deployed as the primary GEMM kernel in our production LLM serving infrastructure. In summary, this paper makes the following contributions.

\begin{itemize}[leftmargin=*]
\item We provide an in-depth analysis of the W4A8 GEMM execution pipeline and identify key performance bottlenecks.
\item We propose \emph{LiquidGEMM}, a high-performance W4A8 GEMM kernel optimized for efficient LLM serving.
\item We develop \emph{LiquidQuant}, a hardware-efficient quantization algorithm that minimizes dequantization overhead on GPUs.
\item We introduce an \emph{implicit fine-grained pipeline} that maximizes hardware utilization through efficient pipeline execution.
\end{itemize}

To evaluate the efficiency of LiquidGEMM, we implement an end-to-end LLM serving system built on top of open-source components including FlashAttention~\cite{dao2023flashattention} for attention computation and PagedAttention~\cite{kwon2023efficient} for KV cache management. Experimental results demonstrate that LiquidGEMM achieves up to 2.90x speedup over the state-of-the-art W4A8 kernel~\cite{lin2024qserve}, and leads to up to 4.94x end-to-end system-level speedup. Compared with various quantized GEMM kernels (W4A16, W8A8, and FP8) in NVIDIA TensorRT-LLM, LiquidGEMM delivers 1.12-1.63x performance gains, and achieves up to 1.63x system-level speedup.

\section{Preliminary} \label{sec:background}

\setlength{\textfloatsep}{0pt}
\begin{figure}[t]\small
    \setlength{\abovecaptionskip}{3pt}
    \setlength{\belowcaptionskip}{0pt}
    \includegraphics[scale=0.72]{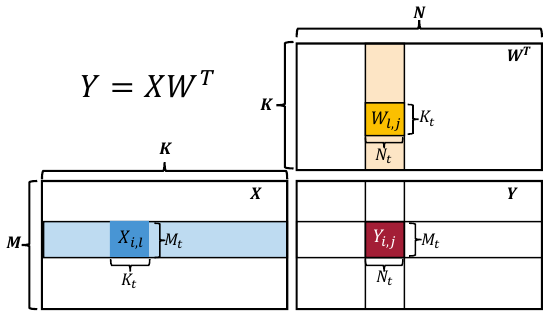}
    \centering
    \caption{Overview of GEMM on GPUs, where $i$, $j$, $l$ denote loop iterations along the $M$, $N$, $K$ dimensions, respectively.}
    \label{fig:gemm_execution}
\end{figure}

\noindent\textbf{Integer Quantization.} This is an important technique to reduce the memory footprint and computational cost of LLMs by converting high-precision floating-point values (\textmd{FP32} or \textmd{FP16}) into low-precision integer representations (e.g., \textmd{INT8} or \textmd{INT4}). This transformation enables more efficient model execution on GPUs that support integer arithmetic. Formally, quantization maps a floating-point tensor $W$ to an $n$-bit integer tensor $Q$ as follows:

\begin{equation}\label{eq:basic_quantization}
    Q = \left\lfloor \frac{W}{s} + z \right\rceil, s = \frac{\max(W) - \min(W)}{\max(Q) - \min(Q)}, z = \left\lfloor \min(Q) - \frac{\min(W)}{s} \right\rceil.
\end{equation}

$s$ is the scaling factor, and $z$ is the zero-point. The operator $\lfloor \cdot \rceil$ denotes rounding to the nearest integer. Since $Q$ is represented using $n$ bits, its dynamic range is constrained to $[0, 2^n - 1]$ for unsigned integers, or $[-2^{n-1}, 2^{n-1}-1]$ for signed integers, depending on the quantization type. The corresponding dequantization process reconstructs an approximate floating-point value $\widehat{W}$ from the quantized integer tensor $Q$:

\begin{equation}\label{eq:basic_dequantization}
    \widehat{W} = (Q - z) \cdot s.
\end{equation}

In practice, two common variants of quantization are used: \emph{asymmetric quantization}, where $z$ is nonzero to accommodate arbitrary input ranges, and \emph{symmetric quantization}, where the range is centered around zero and $z$ is set to 0. In asymmetric quantization, the integer range is given by $\max(Q) - \min(Q) = 2^n - 1$, while in symmetric quantization, the range becomes $2^n - 2$ because $|\max(Q)| = |\min(Q)|$. Compared to symmetric quantization, asymmetric quantization can fully utilize the available value range but requires an additional subtraction operation during dequantization.

\vspace{3pt}
\noindent\textbf{GEMM on GPUs.} Figure~\ref{fig:gemm_execution} provides an overview of GEMM execution on GPUs. Given a GEMM operation $Y = XW^T$, where $X \in \mathbb{R}^{M \times K}$ is the input tensor, $W^T \in \mathbb{R}^{K \times N}$ is the weight matrix, and $Y \in \mathbb{R}^{M \times N}$ is the output, the GPU partitions $Y$ into tiles of size $M_t \times N_t$, each handled by a thread block. To compute its assigned tile, a thread block iterates over the $K$ dimension in steps of $K_t$, performing a sequence of smaller GEMMs of size $M_t \times N_t \times K_t$. In each iteration, it loads the corresponding slices of $X$ and $W$, performs multiply-accumulate operations, and updates the output tile. This iteration over the $K$ dimension, called \emph{main loop}, dominates the overall computation cost of GEMM. Each output tile is further divided into \emph{fragments}, with each warp computing a fragment using \texttt{MMA} (Matrix Multiply-Accumulate) instructions on Tensor Cores. These hardware-accelerated Tensor Cores are optimized for small matrix shapes (e.g., $64 \times 256 \times 32$), allowing high-throughput computation by processing multiple fragments in parallel. For simplicity, we use the terms \emph{tile} and \emph{fragment} interchangeably throughout the paper, as their distinction does not affect the core analysis.

\setlength{\textfloatsep}{0pt}
\begin{figure}[t]\small
    \setlength{\abovecaptionskip}{3pt}
    \setlength{\belowcaptionskip}{0pt}
    \includegraphics[scale=0.72]{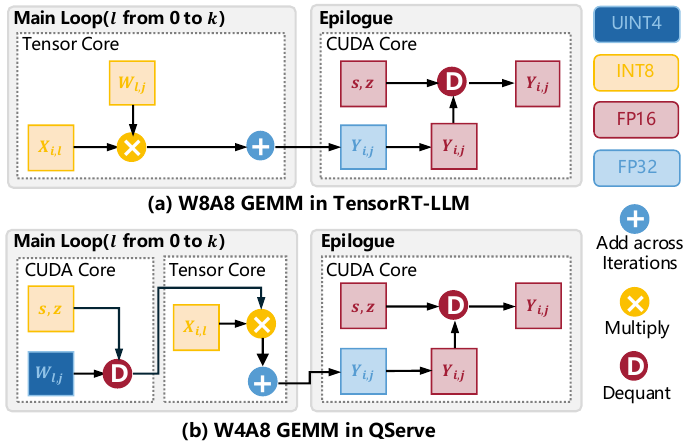}
    \centering
    \caption{Comparison of W8A8 GEMM in TensorRT-LLM and W4A8 GEMM in QServe.}
    \label{fig:W4A8_W8A8}
\end{figure}

Tensor Cores natively support operations on operands with matching symmetric precision, i.e., both weights and activations have the same data type. Based on the precision of the input matrices, GEMM can be categorized into two types: \emph{symmetric GEMM}, where both operands share the same type, and \emph{asymmetric GEMM}, where weights and activations differ in precision (typically, weights have lower bit-width). In asymmetric GEMM, weights must be dequantized during the main loop before being processed by Tensor Cores. Figure~\ref{fig:W4A8_W8A8} compares W4A8, an asymmetric GEMM, with W8A8, a symmetric GEMM. In W4A8, dequantization is performed on CUDA Cores during the main loop prior to MMA on Tensor Cores. In contrast, W8A8 executes the main loop entirely on Tensor Cores, with dequantization deferred to the epilogue.

\section{Motivation} \label{sec:motivation}

We evaluate the practical performance of \textbf{W4A8} GEMM in LLM serving, comparing it with representative quantization methods. Specifically, we benchmark QServe~\cite{lin2024qserve} (\textbf{W4A8}), TRT-W4A16 (\textbf{W4A16}), TRT-W8A8 (\textbf{W8A8}), TRT-FP8 (\textbf{FP8}) and TRT-FP16 (\textbf{FP16}), where TRT refers to TensorRT-LLM~\cite{tensorrt_llm} developed by NVIDIA. We also consider Atom~\cite{zhao2024atom} (\textbf{W4A4}) and QQQ~\cite{zhang2024qqq} (\textbf{W4A8}). However, Atom performs slower on H800 GPUs, as Tensor Cores do not support INT4. QQQ also underperforms compared to QServe. Thus, we omit both Atom and QQQ from further evaluation.

\begin{figure}[t]\small
    \setlength{\abovecaptionskip}{3pt}
    \setlength{\belowcaptionskip}{-10pt}
    \includegraphics[width=1.0\linewidth]{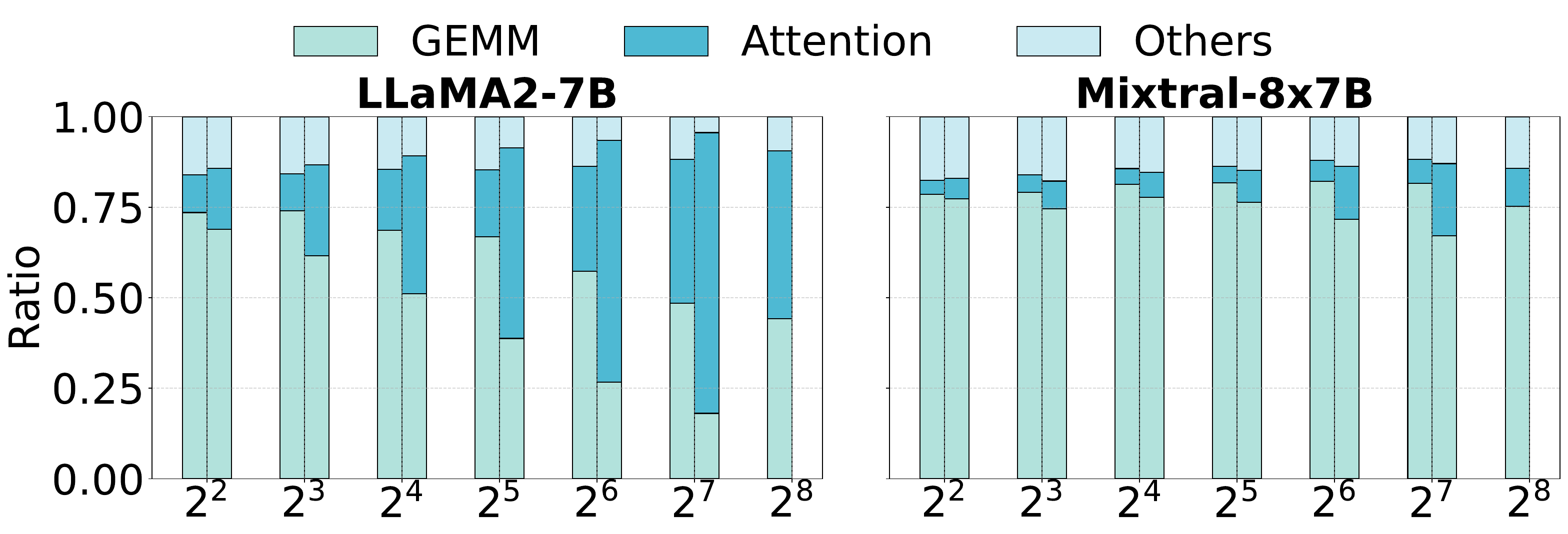}
    \centering
    \caption{Time breakdown of inference for input lengths 128 (left) and 1024 (right). The bar for batch size 256 at length 1024 is omitted due to out-of-memory.}
    \label{fig:end2end_ratio_motivation}
\end{figure}

\begin{figure}[t]
	\setlength{\abovecaptionskip}{0pt}
	\setlength{\belowcaptionskip}{5pt}
		\captionsetup[subfigure]{aboveskip=0pt,belowskip=0pt}
	\centering

    \includegraphics[width=0.9\columnwidth]{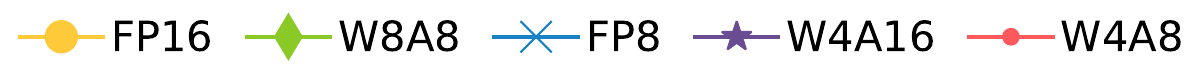}
	\begin{subfigure}[t]{0.23\textwidth}
		\centering
		\includegraphics[width=\textwidth]{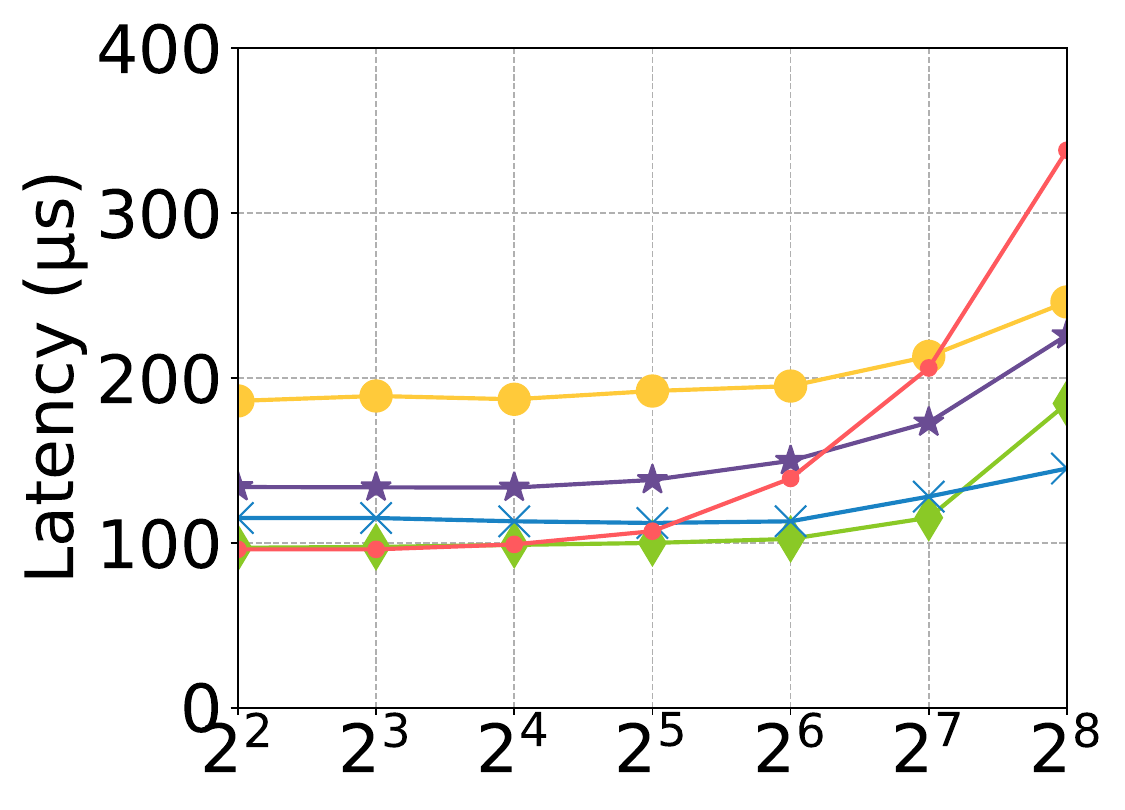}
		\caption{LLaMA2-7B}
		\label{fig:gemm_latency_llama2-7b}
	\end{subfigure}
        \begin{subfigure}[t]{0.23\textwidth}
    		\centering
    		\includegraphics[width=\textwidth]{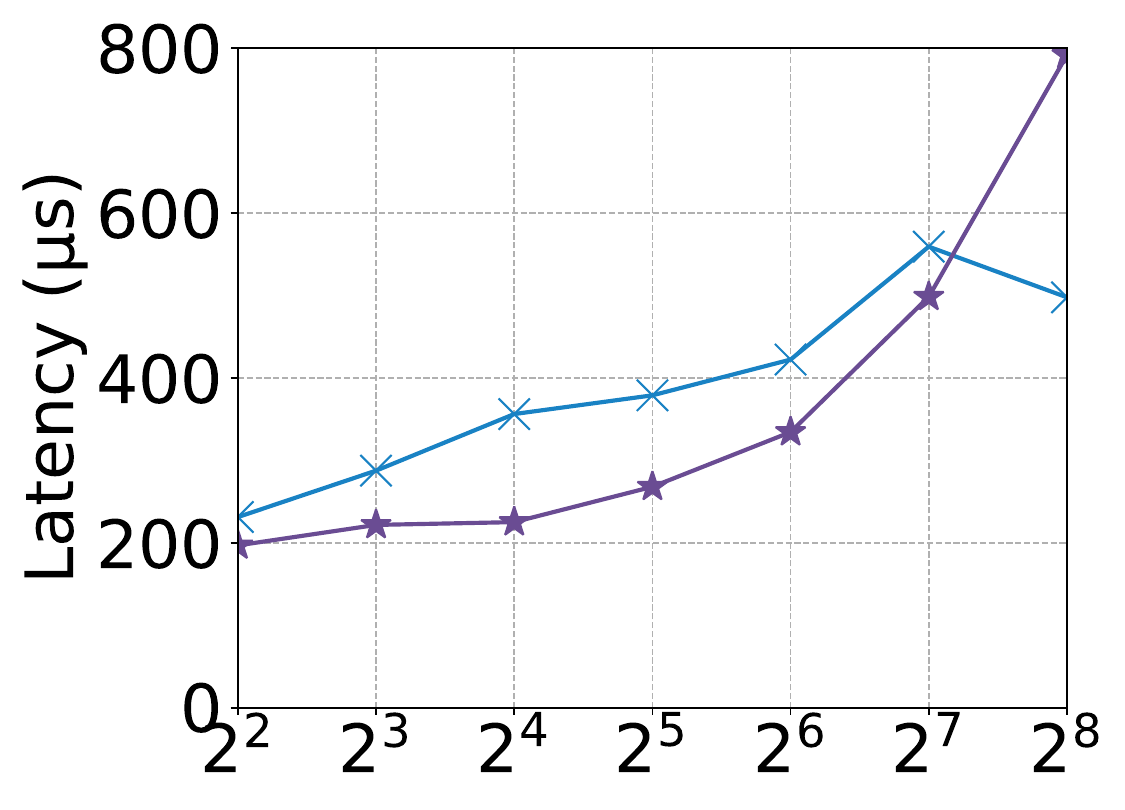}
    		\caption{Mixtral-8x7b}
    		\label{fig:gemm_latency_mixtral-8x7b}
    	\end{subfigure}
	\caption{GEMM latency on a single transformer layer with batch sizes ranging from 4 to 256.}
	\label{fig:motivation_gemm_latency}
\end{figure}

\subsection{Gap between Roofline Analysis and Practice}\label{sec:theory_practice}

To evaluate the practical performance of these GEMM configurations, we benchmark LLM serving on LLaMA2-7B (a dense model) and Mixtral-8×7B (an MoE model) using H800 GPU, with batch sizes ranging from 4 to 256. We select W8A8 quantization for LLaMA2-7B and FP8 quantization for Mixtral-8×7B, as W8A8 quantization does not currently support Mixtral-8×7B. We consider two input-output length settings: 1) 1024 input tokens and 512 output tokens; and 2) 128 input tokens and 128 output tokens. Note that increasing token length does not impact the GEMM workload in the FFN and projection (PROJ) layers during decoding, but it does increase attention computation. Figure~\ref{fig:end2end_ratio_motivation} shows the proportion of GEMM latency (from FFN and PROJ layers) in end-to-end inference. We observe that GEMM dominates latency at small batch sizes and still accounts for over 20\% of total latency at large batch sizes with long sequences on LLaMA2-7B. For Mixtral-8×7B, GEMM remains the primary contributor to latency across all test cases due to the need to run separate GEMMs for each expert. These results highlight the fundamental role of GEMM in LLM serving performance.

Figure~\ref{fig:motivation_gemm_latency} shows the average per-layer GEMM latency during decoding. Contrary to roofline-based predictions, W4A8 performs similarly to W8A8 at small batch sizes ($M \leqslant 64$), but becomes nearly 2x slower at larger batch sizes ($M \geqslant 128$), where it is expected to be competitive. Notably, W4A8 even underperforms FP16 and W4A16, which involve no or partial quantization. Only results for FP8 and W4A16 are reported on Mixtral-8×7B, as other systems lack support for this model. Latency on Mixtral is also substantially higher than on LLaMA2-7B. In summary, despite theoretical expectations that W4A8 should outperform W8A8 in the memory-bound regime and match its performance in the compute-bound regime, our results show that existing W4A8 implementations consistently fall short, particularly under compute-bound conditions, highlighting a clear gap between theoretical potential and practical performance.

\subsection{A Deep Dive into the GEMM Processing}\label{sec:deep_dive}

To understand the performance gap, we first profile the overhead of W4A8 dequantization, then develop a cost model to capture the key performance factors.

\vspace{3pt}
\noindent\textbf{Dequantization Overhead of QServe.} We focus on the main loop of QServe, as the $K$ dimension is typically much larger than the tile size $K_t$ and dominates the overall GEMM cost. In each iteration, QServe dequantizes weights from UINT4 ($Q_{u4}$) to \textmd{INT8} ($Q_{i8}$) using register-level parallelism, with each 32-bit register processing four elements. Given $Q_{i8}$ and $Q_{u4}$, the scale factor $s_{i8}$ and the zero point $z_{i8}$ can be calculated based on Equation~\ref{eq:basic_quantization} offline. To avoid overflow during register-level arithmetic, QServe applies two techniques: 1) \emph{Progressive Quantization:} It constrains $Q_{i8}$ to the range $[-119, 119]$, ensuring $Q_{u4} \cdot s_{i8}$ remains within the valid range; and 2) \emph{Subtraction after Multiplication:} Instead of subtracting $z_{i8}$ before multiplication (as in Equation~\ref{eq:basic_dequantization}), QServe defers subtraction to avoid multiplying negative values, computing $Q_{u4} \cdot s_{i8} - s_{i8} \cdot z_{i8}$.

Despite these efforts, the subtraction step can still overflow. To mitigate this, QServe relies on the \texttt{vadd} instruction to add four 8-bit elements packed into a 32-bit register. However, \texttt{vadd} is not a native hardware instruction and is lowered to a dozen low-level operations, creating significant pressure on CUDA Cores. Due to the large performance gap between CUDA Cores and Tensor Cores (see Figure~\ref{fig:roofline}), this overhead becomes a bottleneck. NVIDIA Nsight profiling on the FFN layer of LLaMA2-7B reveals that subtraction involving \texttt{vadd} accounts for 21\% of warp stalls, highlighting the performance cost of QServe’s dequantization strategy.

\vspace{3pt}
\noindent\textbf{Cost Model.}
We next propose a cost model to capture key performance factors of the pipelined GEMM execution with dequantization described in Section~\ref{sec:background}. Suppose the tile size is $M_t \times N_t \times K_t$.
Then, the number of output tiles is $m \times n$, where
$m = \lceil \tfrac{M}{M_t} \rceil, 
    n = \lceil \tfrac{N}{N_t} \rceil,$
and each tile requires $k = \lceil \tfrac{K}{K_t}\rceil$
iterations to complete the main loop. Each iteration of the main loop consists of two stages: \emph{data loading} and \emph{computation}. We first analyze the cost of a single iteration, and then extend the analysis to the full pipelined execution.

\emph{Data Loading.} The per-iteration data loading time is given by Equation~\ref{eq:data_loading}, where $\phi_{\mathrm{BD}}^{x}$ denotes the \emph{block-level} throughput (in elements/s) for loading data of type $x$, based on the effective memory bandwidth available to a thread block. In LLM serving, the activation matrix is typically small and reused from fast memory, so the cost is dominated by transferring weight from global memory.

\begin{equation} \label{eq:data_loading}
    T_{\mathrm{LD}} 
    = \frac{(M_t \cdot K_t + N_t \cdot K_t)}{\phi_{\mathrm{BD}}^{x}}
    \;\approx\; 
    \frac{N_t \cdot K_t}{\phi_{\mathrm{BD}}^{x}}.
\end{equation}

\emph{Computation.} The computation stage includes: 1) dequantization on CUDA Cores; and 2) MMA on Tensor Cores. Hence, the computation time per iteration is given by:
\begin{equation} \label{eq:computation}
    T_{\mathrm{COMP}}
    = \frac{\alpha \cdot N_t \cdot K_t}{\phi_{\mathrm{CUDA}}}
      \;+\;
      \frac{2 \,\cdot\, \min(M_t,\,M) \,\cdot\, N_t \cdot K_t}{\phi_{\mathrm{TC}}^{y}},
\end{equation}
where $\alpha$ is the number of instructions required to dequantize one weight element, 
$\phi_{\mathrm{CUDA}}$ is the block-level CUDA-Core throughput (OPs/s), and 
$\phi_{\mathrm{TC}}^{y}$ is the block-level Tensor-Core throughput (OPs/s) for data type $y$. One \textmd{MAC} (multiply-accumulate) equals two operations (one multiply and one addition). After all iterations, each output tile must be written back to global memory, incurring an epilogue cost. Since the main loop typically dominates, we omit that epilogue cost.

\emph{Single-Tile Execution.}
The total time $T_{t}$ for a thread block to compute one output tile includes an initial pipeline fill plus repeated overlapped loading and compute. For large $k$, fill and drain overheads are negligible, so $T_{t}$ can be approximated as Equation~\ref{eq:time_cost_one_tile}:

\begin{equation} \label{eq:time_cost_one_tile}
\begin{split}
    T_{t} &= T_{\mathrm{LD}} + T_{\mathrm{COMP}} 
            + (k - 1)\cdot\max(T_{\mathrm{LD}},\,T_{\mathrm{COMP}}) \\
          &\approx k\cdot \max(T_{\mathrm{LD}},\,T_{\mathrm{COMP}}).
\end{split}
\end{equation}

\emph{GPU-Level Execution.}
Assume a device with $S$ streaming multiprocessors, each capable of running up to $L$ thread blocks concurrently.
Denote the \emph{device-level} throughputs by 
$\Phi_{\mathrm{BD}}^{x}$ (memory), $\Phi_{\mathrm{CUDA}}$ (CUDA Cores), and $\Phi_{\mathrm{TC}}^{y}$ (Tensor Cores). 
Since $M_t, N_t, K_t$ are small, we typically have $N \gg N_t$ and $K \gg K_t$; $M$ depends on batch size. 
Given $m \times n$ total tiles, the overall execution time $T$ is approximated by:
\begin{equation} \label{eq:gpu_level_time}
\begin{split}
    T &= \frac{m \cdot n}{S \cdot L} \cdot T_t = m \cdot  \max\biggl(\frac{n \cdot k}{S \cdot L} \cdot T_{\text{LD}}, \frac{n \cdot k}{S \cdot L} \cdot T_{\text{COMP}}\biggr) \\[5pt]
    &\approx m \cdot \max\biggl(\frac{N \cdot K}{\phi_{BD}^{x} \cdot S \cdot L},\; \frac{\alpha \cdot N \cdot K}{\phi_{\text{CUDA}} \cdot S \cdot L} + \frac{2 \cdot \min(M_t, M) \cdot N \cdot K }{\phi_{\text{TC}} ^ {y}\cdot S \cdot L}\biggr) \\[5pt]
    &= \lceil \frac{M}{M_t} \rceil \cdot \max\biggl(\underbrace{\frac{N \cdot K}{\Phi_{\text{BD}} ^ x}}_{T_{LD}},\; \underbrace{\alpha \cdot \frac{N \cdot K}{\Phi_{\text{CUDA}}}}_{T_{DQ}} + \underbrace{\min(M_t, M) \cdot \frac{2 \cdot N \cdot K}{\Phi_{\text{TC}} ^ {y}}}_{T_{MMA}}\biggr),
\end{split}
\end{equation}
where $T_{LD}$, $T_{DQ}$ and $T_{MMA}$ denote the time of data load, dequantization and MMA, respectively. For brevity, we use the same notation $T_{LD}$ to denote the data loading time per iteration. We define the effective output height as $\min(M_t, M)$ to account for cases where the batch size is smaller than the tile size. The cost model highlights how GEMM performance is influenced by batch size $M$, hardware metrics ( $\Phi_{\text{BD}}$, $\Phi_{\text{CUDA}}$, and Tensor Core throughput $\Phi_{\text{TC}}$), and quantization precision (weight bit-width $x$ and activation bit-width $y$). 

\subsection{Insights from Profiling and Analysis}\label{sec:insight}

\noindent\textbf{Root Cause of the Gap.} According to the model, without dequantization overhead, W4A8 and W8A8 should exhibit similar performance in compute-bound scenarios since both use INT8 MMA and share the same $T_{\text{MMA}}$. In memory-bound cases, W4A8 is expected to outperform W8A8 due to its lower memory load ($T_{\text{LD}}$). The transition point occurs when $T_{\text{LD}} = T_{\text{MMA}}$, corresponding to batch size thresholds of 150 for W4A8 and 300 for W8A8 on H100, based on the metrics in Figure~\ref{fig:roofline}. The results are consistent with prior roofline-based analyses~\cite{zhao2024atom,zhang2024qqq}.

However, dequantization shifts this performance curve. The overhead $T_{\text{DQ}}$, determined by the weight matrix size, becomes significant due to the limited compute capacity of CUDA Cores ($\Phi_{\text{CUDA}}$) and high per-element cost $\alpha$ from overflow handling. As a result, W4A8 delivers similar performance to W8A8 in memory-bound cases, despite having a lower $T_{\text{LD}}$, and performs up to 2x slower in compute-bound scenarios, as shown in Section~\ref{sec:theory_practice}. While one might expect to amortize $T_{\text{DQ}}$ by increasing the batch size $M$, the arithmetic intensity is ultimately bounded by the tile size $M_t$, which is constrained by shared memory. This limitation prevents $T_{\text{DQ}}$ from being effectively hidden, resulting in a notable gap between theoretical expectations and observed performance.

\vspace{3pt}
\noindent\textbf{Implication on Efficient GEMM Design.} The cost model suggests two key design principles for efficient W4A8 GEMM. First, weight loading, dequantization, and MMA should be fully pipelined across heterogeneous hardware units (TMA, CUDA Cores, and Tensor Cores) to avoid serialization bottlenecks from dequantization. Second, dequantization must be highly hardware-efficient to enable effective overlap with other stages. In principle, to match the latency of weight loading in memory-bound scenarios ($T_{\text{DQ}} \leqslant T_{\text{LD}}$), the instruction cost per dequantized element must be $\alpha \leqslant 5.07$ on H100, based on metrics in Figure~\ref{fig:roofline}. In compute-bound settings ($T_{\text{DQ}} \leqslant T_{\text{MMA}}$), this threshold becomes $\alpha \leqslant 5.05$ when $M = 150$. Additionally, CUDA Cores must perform auxiliary tasks such as address computation, further increasing the computational load. Together, these constraints underscore the challenge of achieving low-overhead dequantization on modern GPUs.

\vspace{3pt}
\noindent\textbf{Implication on LLM Serving.} We briefly discuss how hardware trends influence LLM serving. In production settings, it is desirable to reach the compute-bound regime at a small batch size to: 1) fully utilize GPU compute capacity; 2) reduce request latency; 3) support long sequences; and 4) minimize operational risks such as hardware faults. Moreover, the batch size is also limited by memory size. However, as shown in Figure~\ref{fig:roofline}, Tensor Core performance is improving faster than memory bandwidth, pushing the memory-to-compute transition point to higher batch sizes, 156 for W8A8 on A100 and 300 on H100, according to our model. In contrast, W4A8 cuts these thresholds in half. This highlights both the value of quantization in enabling efficient inference and the importance of high-performance W4A8 GEMM kernels.

To this end, we propose \textbf{LiquidGEMM}, a hardware-efficient W4A8 GEMM kernel for high-performance LLM serving. In the following sections, we introduce our quantization algorithm, describe the kernel pipeline design and optimization, and present the implementation of an end-to-end LLM serving system for evaluation.

\section{Quantization Algorithm} \label{sec:quantization}

To address dequantization overflow issues, we propose \textbf{LiquidQuant} (LQQ), a hardware-efficient \textbf{W4A8} quantization scheme natively supported by hardware instructions.

\vspace{3pt}
\noindent\textbf{Quantization.} To improve low-bit quantization accuracy, LQQ adopts a group-wise quantization strategy~\cite{zhao2024atom,lin2024qserve,zhang2024qqq,frantar2022gptq,lin2024awq} and a two-level quantization framework that converts \textmd{FP16} weights to \textmd{UINT4}. Since the first-level dequantization occurs in the GEMM epilogue and incurs negligible overhead, our focus is on the second-level quantization. Specifically, following QServe~\cite{lin2024qserve}, the first level quantizes $W$ to an \textmd{INT8} tensor $Q_{i8}$ using per-channel scales, as defined in Equation~\ref{eq:basic_quantization}. We also adopt the \emph{protective quantization range} in Section~\ref{sec:deep_dive}, which restricts $Q_{i8} \in [-119, 119]$ to prevent overflow during scaling in dequantization (see~\cite{lin2024qserve} for proof).

The second-level converts \textmd{INT8} to \textmd{UINT4}. Our key idea is to shift the symmetric range of $Q_{i8}$ into the unsigned domain of a \textmd{UINT8} tensor $Q_{u8}$, and then quantize $Q_{u8}$ to $Q_{u4}$. This design aligns with our dequantization method to eliminate potential overflow during inference, as we will prove at the end of this section. The quantization process is defined in Equation~\ref{eq:int8_uint4}. We omit the zero point $z_{u8}$, as both $\min(Q_{u8})$ and $\min(Q_{u4})$ are zero.

\begin{equation}\label{eq:int8_uint4}
    Q_{u8} = Q_{i8} - \min(Q_{i8}), \quad
    Q_{u4} = \left\lfloor \frac{Q_{u8}}{s_{u8}} \right\rceil, \quad
    s_{u8} = \frac{\max(Q_{u8})}{\max(Q_{u4})}.
\end{equation}

Compared to the standard quantization in Equation~\ref{eq:basic_quantization}, our method introduces a simple shift from $Q_{i8}$ to $Q_{u8}$, performed entirely offline. The core optimization focuses on the online dequantization, which is crucial for efficient LLM serving.

\vspace{3pt}
\noindent\textbf{Dequantization.} Based on Equation~\ref{eq:int8_uint4}, we dequantize the tensor from \textmd{UINT4} back to \textmd{INT8} during inference as follows:

\begin{equation} \label{eq:uint4_int8}
    \widehat{Q}_{i8} = \widehat{Q}_{u8} + \min(Q_{i8}) = Q_{u4} \cdot s_{u8} + \min(Q_{i8}).
\end{equation}

To ensure no overflow, we must guarantee that this computation remains within valid numeric ranges. From Equation~\ref{eq:int8_uint4}, we know the scale factor satisfies $s_{u8} \leqslant \lfloor \frac{119 - (-119)}{15} \rceil = 16$. Since $Q_{u4} \in [0, 15]$, we have $\widehat{Q}_{u8} = Q_{u4} \cdot s_{u8} \leqslant 15 \times 16 = 240$, which stays within the \textmd{UINT8} range, avoiding overflow during multiplication.

However, directly adding $\min(Q_{i8})$, which can be negative, can lead to wraparound issues. We illustrate this with an example. Suppose $Q_{u4} = 15$, $\max(Q_{i8}) = 119$, and $\min(Q_{i8}) = -104$. Then, we have $s_{u8} = \lfloor \frac{119 - (-104)}{15} \rceil = \lfloor \frac{223}{15} \rceil = 15$, and the expected result is:
$\widehat{Q}_{i8} = Q_{u4} \cdot s_{u8} + \min(Q_{i8}) = 15 \times 15 + (-104) = 225 - 104 = 121.$
In binary, $Q_{u8} = 225$ is represented as \texttt{1110 0001}, and $\min(Q_{i8}) = -104$ is represented as \texttt{1001 1000} in two's complement form. If the addition is performed at the bit level without type promotion, \texttt{1110 0001} + \texttt{1001 1000} = \texttt{1 0111 1001}, which is overflow. Alternatively, casting $Q_{u8}$ to \textmd{INT8} before the addition is also invalid, since \texttt{1110 0001} represents $-31$ in \textmd{INT8}, not 225. This example highlights that the addition step requires careful handling beyond standard hardware instructions.

LQQ introduces a sweet dequantization method, combined with the shifted quantization, to eliminate overflow by using properties of two's complement representation: an \textmd{INT8} value $i$ and a \textmd{UINT8} value $j$ share the same binary representation if $i \equiv j \pmod{2^8}$. For example, $-3 \equiv 253 \pmod{2^8}$, and both are represented as \texttt{1111 1101}. Using this property, we rewrite Equation~\ref{eq:uint4_int8} as: \begin{equation} \label{eq:dequant_adjusted}
\begin{split}
    \widehat{Q}_{i8} &\equiv Q_{u4} \cdot s_{u8} + \min(Q_{i8}) + x \cdot 2^{8} \pmod{2^8} \\
    &\equiv Q_{u4} \cdot s_{u8} + (2^7 + \min(Q_{i8})) + (2x - 1) \cdot 2^{7} \pmod{2^8},
\end{split}
\end{equation} where $x$ is an integer. We next prove that the computation in Equation~\ref{eq:dequant_adjusted} avoids overflow, i.e., all intermediate results remain within the \textmd{UINT8} range, by properly controlling the value of $x$.

\begin{proof}
    Let $q_i$ be an element in $Q_{i8}$ after the first-level quantization, and let $q_u = q_i - \min(Q_{i8})$ be the corresponding element in $Q_{u8}$. According to Equation~\ref{eq:dequant_adjusted}, the dequantized computation process can be expressed as:

\begin{equation} \label{eq:offset_rotation}
    \widehat{q}_{i} \equiv 
    \underbrace{\left\lfloor \frac{q_u}{s_{u8}} \right\rceil \cdot s_{u8}}_{\widehat{q}_{u} \in [0, 255]} + 
    \underbrace{(2^7 + \min(Q_{i8}))}_{a \in [0, 255]} + 
    \underbrace{(2x - 1) \cdot 2^7}_{b} 
    \pmod{2^8}.
\end{equation}

We first show that $\widehat{q}_{u} + a$ is bounded within \textmd{UINT8}. Since $s_{u8} \leqslant 16$ and $q_u \leqslant \max(Q_{i8}) - \min(Q_{i8}) = 238$, we have:

\begin{equation} \label{eq:no_overflow1}
    \begin{split}
    \widehat{q}_{u} + a &= \lfloor \frac{q_u}{s_{u8}}\rceil \cdot s_{u8} + a \leqslant q_{u} + \frac{s_{u8}}{2} + a \\
    &\leqslant (\max(Q_{i8}) - \min(Q_{i8})) + 8 + (2^7 + \min(Q_{i8}))\\
    &= \max(Q_{i8}) + 8 + 2^7 \leqslant 119 + 8 + 128 = 255.
    \end{split}
\end{equation}

Next, to ensure the final result $\widehat{q}_{u} + a + b$ also stays within $[0, 255]$, we control the value of $x$ as follows: if $\widehat{q}_{u} + a \geqslant 128$, set $x = 0$ so $b = -128$; otherwise, set $x = 1$ so $b = 128$. This guarantees that the computation in Equation~\ref{eq:offset_rotation} is overflow-free within \textmd{UINT8}.
\end{proof}

\vspace{3pt}
\noindent\textbf{Hardware-Efficient Computation.}  
Checking $\widehat{q}_{u} + a$ and determining $x$ at runtime can introduce significant overhead in the main-loop of GEMM. Upon analysis, we observe that adding $b$ is equivalent to flipping the \emph{most significant bit} of $\widehat{q}_{u} + a$. Therefore, dequantization can be performed as:
\begin{equation} \label{eq:lqq_dequant}
   \widehat{Q}_{i8} = (Q_{u4} \cdot s_{u8} + a) \oplus 0\text{x}80,
\end{equation}
where $a = 2^7 + \min(Q_{i8})$ is precomputed offline and $\oplus$ denotes the \texttt{XOR} operation. This formulation keeps all intermediate values within the \textmd{UINT8} range, avoids overflow, and enables efficient hardware execution (see Section \ref{sec:hardware_dequantization}). For the first-level dequantization, LQQ follows the standard process in Equation \ref{eq:basic_dequantization}.
\section{High Performance \textbf{W4A8} GEMM Kernel} \label{sec:mpGEMM}

Building on LiquidQuant (LQQ), we propose LiquidGEMM, a high-performance \textbf{W4A8} GEMM kernel featuring an asynchronous computation pipeline. We use the current cloud workhorse GPU, the H800, to illustrate the kernel. To optimize execution, we compute $Y = (WX^T)^T$ instead of $Y = XW^T$, as explained in Section~\ref{sec:other_optimization}.

\subsection{Design of Async Computation Pipeline}

\setlength{\textfloatsep}{0pt}
\begin{figure}[t]\small
    \setlength{\abovecaptionskip}{3pt}
    \setlength{\belowcaptionskip}{0pt}
    \includegraphics[scale=0.62]{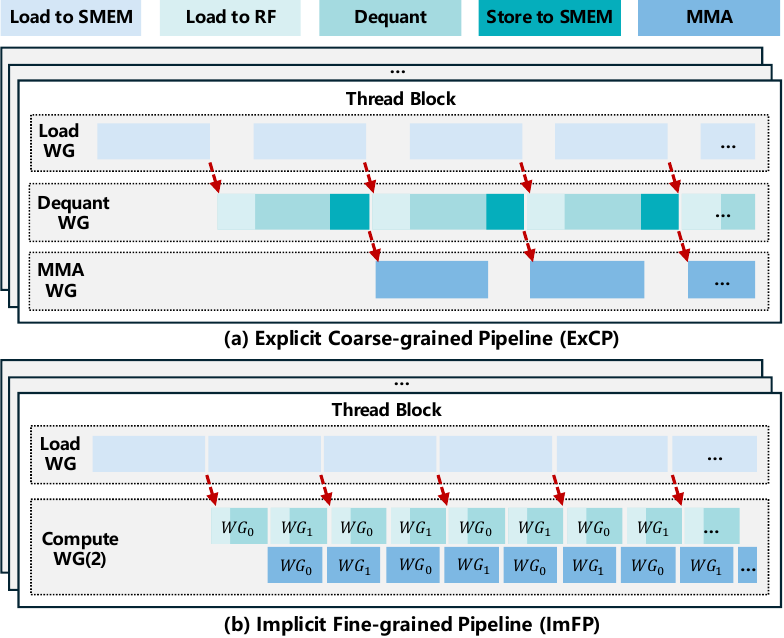}
    \centering
    \caption{Comparison of Explicit Coarse-Grained Pipeline (ExCP) and Implicit Fine-Grained Pipeline (ImFP) designs.}
    \label{fig:pipeline}
\end{figure}

\noindent\textbf{Explicit Coarse-Grained Pipeline (ExCP).}  
High-performance GEMM libraries such as CUTLASS use \emph{warp specialization} to overlap weight loading and computation. In this model, warps within a thread block are divided into specialized roles, such as \emph{Load Warps} and \emph{MMA Warps}, which operate asynchronously in a producer-consumer fashion. On H800, warps are grouped into \emph{warp groups} (WGs), with each group consisting of four warps (128 threads) that work collectively. A straightforward idea in dequantization context is to apply it to W4A8 computation. Specifically, as shown in Figure \ref{fig:pipeline}, we design a three-stage pipeline in which three WGs are assigned to load weights, perform dequantization, and execute MMA, respectively. Each stage is mapped to a distinct hardware unit: weight loading via TMA, dequantization via CUDA Cores, and MMA via Tensor Cores. These stages operate concurrently, enabling overlap of $T_{\text{LD}}$, $T_{\text{DQ}}$, and $T_{\text{MMA}}$. We refer to this approach as the \emph{explicit coarse-grained pipeline} (ExCP).

However, ExCP can introduce pipeline bubbles that degrade GEMM efficiency due to its coarse-grained execution and explicit scheduling of warp groups. In particular, the Dequant WG loads weights from SMEM, previously loaded by the Load WG from GMEM, into RF for dequantization on CUDA Cores. After dequantization, it writes the weights back to SMEM and signals the MMA WG to begin execution. This round-trip data movement between RF and SMEM incurs non-trivial overhead and increases the workload of the Dequant WG, creating pipeline stalls. Moreover, software-based synchronization between the Dequant and MMA WGs adds further overhead.

\vspace{3pt}
\noindent\textbf{Implicit Fine-Grained Pipeline (ImFP).}  
To address the inefficiencies of ExCP, we propose the \emph{implicit fine-grained pipeline} (ImFP). Unlike ExCP, which assigns separate WGs for dequantization and MMA, ImFP uses a unified \emph{Compute WG} responsible for both tasks. This eliminates the need to write dequantized results from RF back to SMEM, reducing data movement overhead (Figure~\ref{fig:pipeline}). To overlap dequantization and MMA, we leverage pipeline stages across different Compute WGs. Specifically, ImFP adopts a fine-grained pipeline using a single-producer, multiple-consumer model. The Load WG acts as the producer, loading weights from GMEM to SMEM and splitting them into fine-grained tasks, each of which is a fragment of the weight matrix. These tasks are then dynamically fetched and processed by multiple Compute WGs, each performing both dequantization and MMA. Since different computation WGs operate on different tasks, dequantization in one WG naturally overlaps with MMA in another, achieving implicit parallelism without explicit synchronization. In our implementation, each thread block consists of one Load WG and two Compute WGs, which effectively balances hardware utilization and task throughput. Experimental results show that ImFP significantly outperforms the coarse-grained ExCP design. Next, we introduce the data loading and computation in detail.

\subsection{Memory Layout and Data Loading}

\setlength{\textfloatsep}{0pt}
\begin{figure}[t]\small

    \setlength{\abovecaptionskip}{3pt}
    \setlength{\belowcaptionskip}{0pt}
    \includegraphics[scale=0.62]{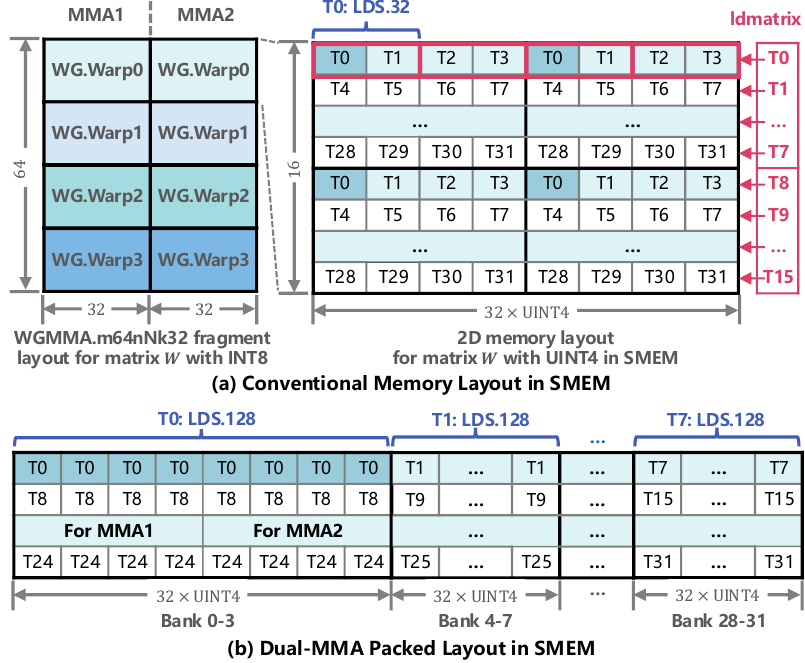}
    \centering
	\caption{Comparison of conventional memory layout and our Dual-MMA packed layout.}
     \label{fig:data_path}
\end{figure}

In each main-loop iteration, the required weight tile is loaded from GMEM to SMEM by the Load WG, and then into RF for dequantization and MMA by the Compute WGs. MMA on Tensor Cores requires a structured data layout across threads to comply with hardware intrinsic requirements. To meet this demand, the memory layout of the weight matrix is crucial, as it directly affects the efficiency of data loading.

\vspace{3pt}
\noindent\textbf{Conventional Approach.}
Modern GPUs support MMA operations on fixed matrix shapes defined by hardware. For \textmd{INT8} inputs, the H800 provides instructions like \texttt{WGMMA.m64nNk32} and \texttt{WGMMA.m64nNk64}, with $N$ ranging from 8 to 256. As illustrated in Figure~\ref{fig:data_path}a, \texttt{WGMMA.m64nNk32} performs a $64 \times N \times 32$ MMA on Tensor Cores, requiring a $64 \times 32$ fragment from matrix $W$. Each warp in a WG loads a $16 \times 32$ segment, with every thread fetching 16 elements into registers using a strided layout: four contiguous elements per group, spaced to match the intrinsic’s tiling pattern. The elements accessed by thread T0 are shown in dark blue in Figure~\ref{fig:data_path}a. To load from SMEM to RF, H800 offers the \texttt{ldmatrix} instruction. Each thread loads 16 contiguous bytes in one transaction and scatters every 4-byte group to the appropriate thread—assuming each element is 1 byte. This assumption fails for W4A8, where elements are compressed to 4 bits. As a result, \texttt{ldmatrix} incorrectly scatters data, e.g., elements meant for T2 and T3 may be delivered to T1, as shown in Figure~\ref{fig:data_path}a. One alternative is to use the \texttt{LDS.32} instruction, which loads 32 bits from a specified address. However, each thread needs only four 4-bit values, meaning half the data is unused, cutting effective bandwidth. Moreover, this approach requires more load instructions and additional address calculations, increasing arithmetic overhead and placing extra burden on CUDA Cores~\cite{lin2024qserve}.

\vspace{3pt}
\noindent\textbf{Dual-MMA Packed Layout.} Inspired by the compute-aware weight reordering in QServe~\cite{lin2024qserve}, we propose the \emph{dual-MMA packed layout} to solve the problem. In a single MMA operation, each thread requires 16 \textmd{UINT4} elements, whereas the coarse-grained \texttt{LDS.128} instruction loads 32 elements per transaction. To exploit this gap, we pack the data required for two consecutive MMA operations per thread and store them contiguously, as shown in Figure~\ref{fig:data_path}b. This enables each thread to load all 32 \textmd{UINT4} elements using a single \texttt{LDS.128} instruction. To satisfy the WGMMA fragment layout, we reorder the weights so that the elements required by each thread across two MMAs are adjacent in memory. Different from QServe storing weights in 2D layout, we arrange these elements in a 1D layout to eliminate shared memory bank conflicts and remove the need for swizzling or complex data packing. This layout supports eight simultaneous \texttt{LDS.128} operations across threads, fully leveraging the shared memory bandwidth. Moreover, the dual-MMA packed layout significantly reduces the number of load instructions and minimizes address computation overhead on CUDA Cores. The weight matrix in GMEM follows the same layout as in SMEM, enabling efficient transfers using \texttt{LDG.128}, the most coarse-grained load instruction available per warp. Since the layout transformation is applied offline, it introduces no runtime overhead.

\subsection{Hardware-Efficient Dequantization}\label{sec:hardware_dequantization}

\setlength{\textfloatsep}{0pt}
\begin{figure}[t]\small
    \setlength{\abovecaptionskip}{3pt}
    \setlength{\belowcaptionskip}{0pt}
    \includegraphics[scale=0.65]{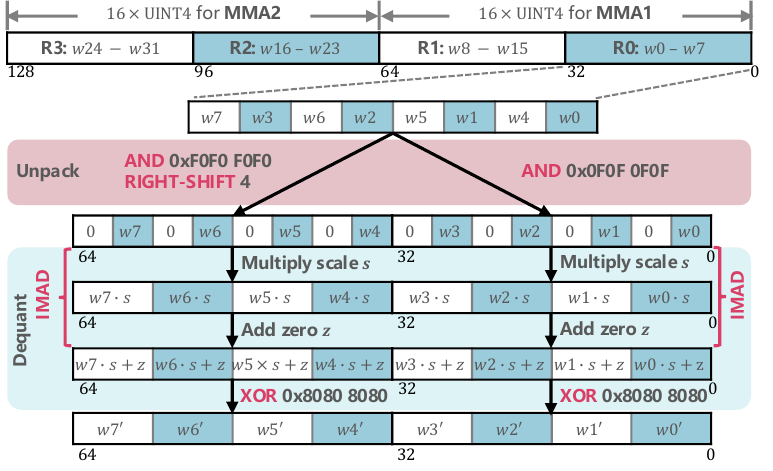}
    \centering
    \caption{Dequantization process using bitwise and IMAD instructions, natively supported by hardware. $s = s_{u8}$ and $z = a$ are calculated offline based on Equation \ref{eq:lqq_dequant}.}
    \label{fig:dequant}
\end{figure}

After loading weights from SMEM to RF, each thread holds 32 \textmd{UINT4} elements packed into four 32-bit registers, as illustrated in Figure~\ref{fig:dequant}. Elements $w8 - w15$ correspond to the first MMA operation, and $w16 - w31$ to the second. We dequantize these weights from \textmd{UINT4} to \textmd{INT8} on CUDA Cores using LQQ (Section~\ref{sec:quantization}).

Figure~\ref{fig:dequant} illustrates the dequantization process. We first apply the unpacking method from QServe~\cite{lin2024qserve} to expand eight 4-bit elements from one register into two registers holding 8-bit values. We then perform dequantization using Equation~\ref{eq:lqq_dequant}: multiplying by the scale factor $s_{u8}$, adding the offset $a$, and applying a final \texttt{XOR}. Because LQQ ensures no overflow, all operations can be executed using native 32-bit hardware instructions, specifically, \texttt{IMAD} for multiply-add and \texttt{XOR} for offset correction. Note that both $s_{u8}$ and $a$ can be precomputed offline. After dequantization, the resulting \textmd{UINT8} elements share the same binary representation as the target \textmd{INT8} values, making them directly usable for subsequent MMA operations on Tensor Cores.

In summary, our method dequantizes four elements using just two hardware arithmetic instructions. Including the unpacking step, eight elements are dequantized with only seven instructions, significantly reducing computational overhead on CUDA Cores, well below the threshold required for effective overlap with weight loading and MMA (Section~\ref{sec:insight}). The first-level dequantization is fused into the GEMM epilogue and incurs negligible cost.

\subsection{Other GEMM Optimizations}\label{sec:other_optimization}

As discussed above, GPU MMA instructions are constrained to fixed matrix shapes defined by hardware. For \textmd{INT8}, H800 fixes the $m$ dimension to 64, while $n$ can vary from 8 to 256 across several configurations. To better utilize Tensor Cores under small batch sizes, we apply a hardware-specific optimization by rewriting $Y = XW^T$ as $Y = (WX^T)^T$, allowing us to choose WGMMA instructions based on the batch size and maximize compute efficiency. Additionally, we adopt standard GEMM optimizations such as persistent kernels. As these techniques are widely used, we omit the details for brevity.

Leveraging the programming primitives of CUTLASS and Cute, we integrate and adapt components such as the tile scheduler, mainloop, and epilogue into a warp-specialized ping-pong kernel. Specifically, our dequantization algorithm is fused into the MMA mainloop, and a Dual-MMA packed layout is used during data loading. We implement WGMMA instructions, barrier synchronization, and general components like TMA in PTX, wrapped by CUTLASS. In contrast, the dequantization logic is implemented directly in CUDA.
\section{LLM Serving System and Offline Quantization} \label{sec:llm_serving_system}

To support end-to-end performance evaluation, we implement an LLM serving system by integrating open-source techniques for key system-level components, including attention computation, KV cache management, and quantization schemes. This section briefly outlines their implementation, along with the offline quantization.

\setlength{\textfloatsep}{0pt}
\begin{figure}[t]\small
    \setlength{\abovecaptionskip}{3pt}
    \setlength{\belowcaptionskip}{0pt}
    \includegraphics[scale=0.62]{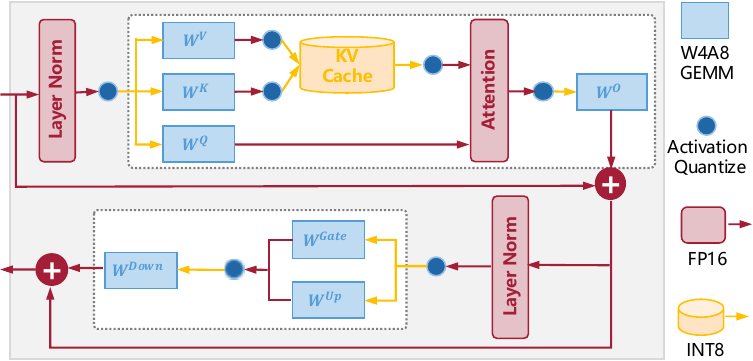}
    \centering
    \caption{Overview of dataflows in our LLM serving system for LLaMA models.}
    \label{fig:dataflow}
\end{figure}

\begin{table*}[t]
\centering
\caption{Peak token generation throughput (tokens/s) of LiquidServe, QServe, and TRT on H800 with 80 GB memory constraint. The number in parentheses indicates the batch size at which peak throughput is achieved. Speedup is reported relative to the best-performing baseline between QServe and TRT. LiquidServe/wo uses the W4A8 GEMM kernel from QServe.}
\label{tab:max_throughput_result}
\resizebox{\textwidth}{!}{%
\begin{tabular}{ccccccccc}
\toprule
\multirow{2}{*}{\textbf{System}} & \textbf{LLaMA1}     & \multicolumn{3}{c}{\textbf{LLaMA2}}                                & \textbf{LLaMA3}       & \textbf{Mistral}      & \textbf{Yi}          & \textbf{Mixtral}      \\ \cline{2-9} 
                                 & \textbf{30B}        & \textbf{7B}          & \textbf{13B}         & \textbf{70B}         & \textbf{8B}           & \textbf{7B}           & \textbf{34B}         & \textbf{8×7B}         \\ \hline
\textbf{TRT-FP16}                & 410 (13)            & 5,521 (128)& 2,701 (64)& OOM                  & 13,920 (256)          & 14,573 (256)          & 1,931 (64)& OOM                   \\ \hline
\textbf{TRT-W4A16}               & 1,170 (48)          & 4,953 (128)& 2,906 (109)          & 2,266 (128)& 12,997 (256)          & 13,513 (256)          & 4,645 (256)          & 5,712 (256)           \\ \hline
\textbf{TRT-W8A8}                & 1,006 (36)          & 5,083 (128)& 2,922 (100)          & 1,166 (46)           & 13,012 (256)          & 13,636 (256)          & 3,860 (128)& NA                    \\ \hline
\textbf{TRT-FP8}                 & 986 (36)            & 5,913 (144)& 3,402 (96)& 948 (45)             & \textbf{16,820 (256)} & \textbf{17,433 (256)} & 4,206 (225)          & 8,296 (256)           \\ \hline
\textbf{QServe}                  & 1,478 (64)          & 5,402 (128)          & 3,311 (124)          & 871 (64)             & 5,240 (128)           & 5,361 (124)           & 1,415 (64)           & NA                    \\ \hline
\textbf{LiquidServe/wo}          & 1,309               & 5,926                & 3,299                & 1,869                & 10,956                & 11,091                & 3,699                & 6,135                 \\ \hline
\textbf{LiquidServe}             & \textbf{1,607 (53)} & \textbf{6,721 (194)} & \textbf{4,105 (119)} & \textbf{3,695 (184)} & 16,694 (256)          & 17,011 (256)          & \textbf{6,999 (256)} & \textbf{10,745 (256)} \\ \hline
\textbf{Speedup}                 & 1.09x                & 1.14x                 & 1.21x& 1.63x& 0.99x                  & 0.98x                  & 1.51x                 & 1.30x                  \\ \bottomrule 
\end{tabular}%
}
\end{table*}

\vspace{3pt}
\noindent\textbf{Serving System.} Figure~\ref{fig:dataflow} illustrates the dataflow of our LLM serving system for LLaMA models. Query, Key, Value, Output, and FFN layers are executed using our proposed LiquidGEMM with W4A8 quantization on weights and activations, producing FP16 outputs. Following TensorRT-LLM~\cite{tensorrt_llm}, KV caches are quantized to INT8 using per-channel static quantization, with scale factors computed offline. To improve memory efficiency, we adopt PagedAttention~\cite{kwon2023efficient} for KV cache management and use FlashAttention-2~\cite{dao2023flashattention} for runtime attention computation. We do not adopt FlashAttention-3~\cite{shah2024flashattention}, as it is tailored for FP8. For activation quantization, we follow SmoothQuant~\cite{xiao2023smoothquant}, dynamically mapping FP16 activations to INT8 on-the-fly via per-token quantization after dividing by the smooth scale. As activation tensors have small memory footprints and low computational overhead, quantization is lightweight and typically fused into other kernels.

\vspace{3pt}
\noindent\textbf{Offline Quantization.} We adopt the SmoothQuant~\cite{xiao2023smoothquant} post-training quantization method to quantize weights offline. Specifically, weights are first scaled by a smooth factor and then quantized using the two-level approach described in Section~\ref{sec:quantization}: per-channel quantization from \textmd{FP16} to \textmd{INT8}, followed by per-group quantization to \textmd{UINT4}. Following OutlierSuppression+~\cite{wei2023outlier}, we apply a grid search to determine the optimal smooth scale. Note that our focus is on optimizing the efficiency of W4A8 GEMM; our method is orthogonal to techniques that improve quantization accuracy and can be seamlessly integrated with such approaches.

\section{Experiments} \label{sec:experiments}

\subsection{Experimental Setup} \label{sec:experiment_setup}

\noindent\textbf{Systems Under Study.} Our W4A8 kernel, \textbf{LiquidGEMM}, is implemented using CUDA and PTX. We refer to the full LLM serving system described in Section~\ref{sec:llm_serving_system} as \textbf{LiquidServe}. By default, weights are quantized using group-wise quantization with a group size of 64, and the KV cache is quantized to \textmd{INT8} using per-channel static quantization. We compare LiquidServe against two baseline systems. The first is \textbf{QServe}~\cite{lin2024qserve}, a state-of-the-art W4A8 LLM serving system featuring an efficient W4A8 GEMM implementation. QServe uses group-wise weight quantization with a group size of 128 by default and quantizes the KV cache to 4-bit. We use the publicly available implementation from GitHub\footnote{https://github.com/mit-han-lab/omniserve, commit hash: 5106921}. The second baseline is \textbf{TensorRT-LLM}~\cite{tensorrt_llm}, an LLM inference framework provided by NVIDIA. We use version 0.16.0 of the implementation\footnote{https://github.com/NVIDIA/TensorRT-LLM, commit hash: 42a7b09}. We include it for comparison under several common precision settings: FP16 (denoted as \textbf{TRT-FP16}), W4A16 (\textbf{TRT-W4A16}), W8A8 (\textbf{TRT-W8A8}), and FP8 (\textbf{TRT-FP8}). The KV caches are quantized per-channel to FP8 for W4A16, FP8, and FP16 configurations, and to INT8 for W8A8.

\vspace{3pt}
\noindent\textbf{Testbed.} We conduct experiments on a Linux server equipped with an Intel Xeon Platinum 8457C CPU, 2.9 TB RAM, and an NVIDIA H800 GPU with 80 GB memory, via cloud access. All system-level evaluations are run under PyTorch 2.4.0 and CUDA 12.4. To isolate and fairly compare GEMM kernel performance, we extract the GEMM kernels from each system and benchmark them using a unified CUDA-based framework\footnote{An internal benchmarking tool used to evaluate GPU kernel performance before deployment.}. This framework supports flexible matrix shape configuration to simulate various model scenarios.

\vspace{3pt}
\noindent\textbf{Experiment Roadmap.} Our evaluation consists of two parts. First, we measure system-level throughput and latency to understand the end-to-end impact of quantization on LLM serving. While our focus is on accelerating W4A8 GEMM, this step provides context on how GEMM efficiency translates to overall system performance. However, system-level performance is also influenced by other factors such as attention computation and KV cache management, which differ across implementations and are outside the scope of this paper. Therefore, we complement this with a second set of experiments that directly benchmark the GEMM kernels in isolation using our unified framework. This enables fair, accurate comparisons under consistent and controlled conditions.

Although our LQQ quantization algorithm is designed to improve efficiency, we also evaluate its impact on model accuracy. We test on LLaMA~\cite{touvron2023llama,touvron2023llama2,grattafiori2024llama}, Mistral-7B~\cite{jiang2023mistral7b}, Mixtral-8×7B~\cite{jiang2024mixtral}, and Yi-34B~\cite{young2024yi}, using perplexity on WikiText2~\cite{wikitext} and zero-shot accuracy on PIQA~\cite{bisk2020piqa}, ARC~\cite{arc}, HellaSwag~\cite{zellers2019hellaswag}, and WinoGrande~\cite{sakaguchi2021winogrande}. Results show that LQQ preserves accuracy. Due to space constraints, detailed results will be released in a full-version technical report.

\subsection{Efficiency Comparison of LLM Serving}

\textbf{Throughput Under Memory Constraint.} We compare the maximum achievable throughput of all systems under the same memory budget (80 GB on an H800 GPU). Following prior work~\cite{lin2024qserve}, we fix the input and output sequence lengths to 1024 and 512, respectively. We vary the batch size from 1 to 256 (or until the system runs out of memory) to identify the optimal configuration, and report the peak throughput achieved by each system.

Table~\ref{tab:max_throughput_result} summarizes the peak throughput across all systems. QServe typically reaches peak performance at batch sizes of 64 or 128, whereas LiquidServe continues to scale with increasing batch size. As a result, QServe outperforms TRT on LLaMA-30B and LLaMA2-13B due to its use of low-bit KV cache, which allows for larger batch sizes, but performs significantly worse on other models. LiquidServe performs slightly below TRT-FP8 on LLaMA3-8B and Mistral-7B, as TRT-FP8 leverages attention kernels optimized for FP8 on H800. However, LiquidServe consistently outperforms all other systems in the remaining cases. The performance advantage is especially pronounced on larger models. For instance, on LLaMA2-70B, LiquidServe achieves a 3.16x speedup over TRT-W8A8 by supporting larger batch sizes through 4-bit weight quantization, and a 1.63x speedup over TRT-W4A16 due to its higher compute throughput with INT8 MMA. These results highlight the practical efficiency gains of W4A8 quantization on the system level.

Nevertheless, as discussed, system performance is influenced by multiple factors such as attention computation and KV cache management. To isolate the contribution of our W4A8 GEMM, we replace LiquidGEMM in LiquidServe with QServe’s W4A8 kernel, denoted as \textbf{LiquidServe/wo}. As shown, LiquidServe achieves a 1.13-1.98x end-to-end speedup over LiquidServe/wo, demonstrating both the critical role of GEMM and the effectiveness of LiquidGEMM.

\begin{figure}[t]\small
    \setlength{\abovecaptionskip}{3pt}
    \setlength{\belowcaptionskip}{0pt}
    \includegraphics[width=1.0\linewidth]{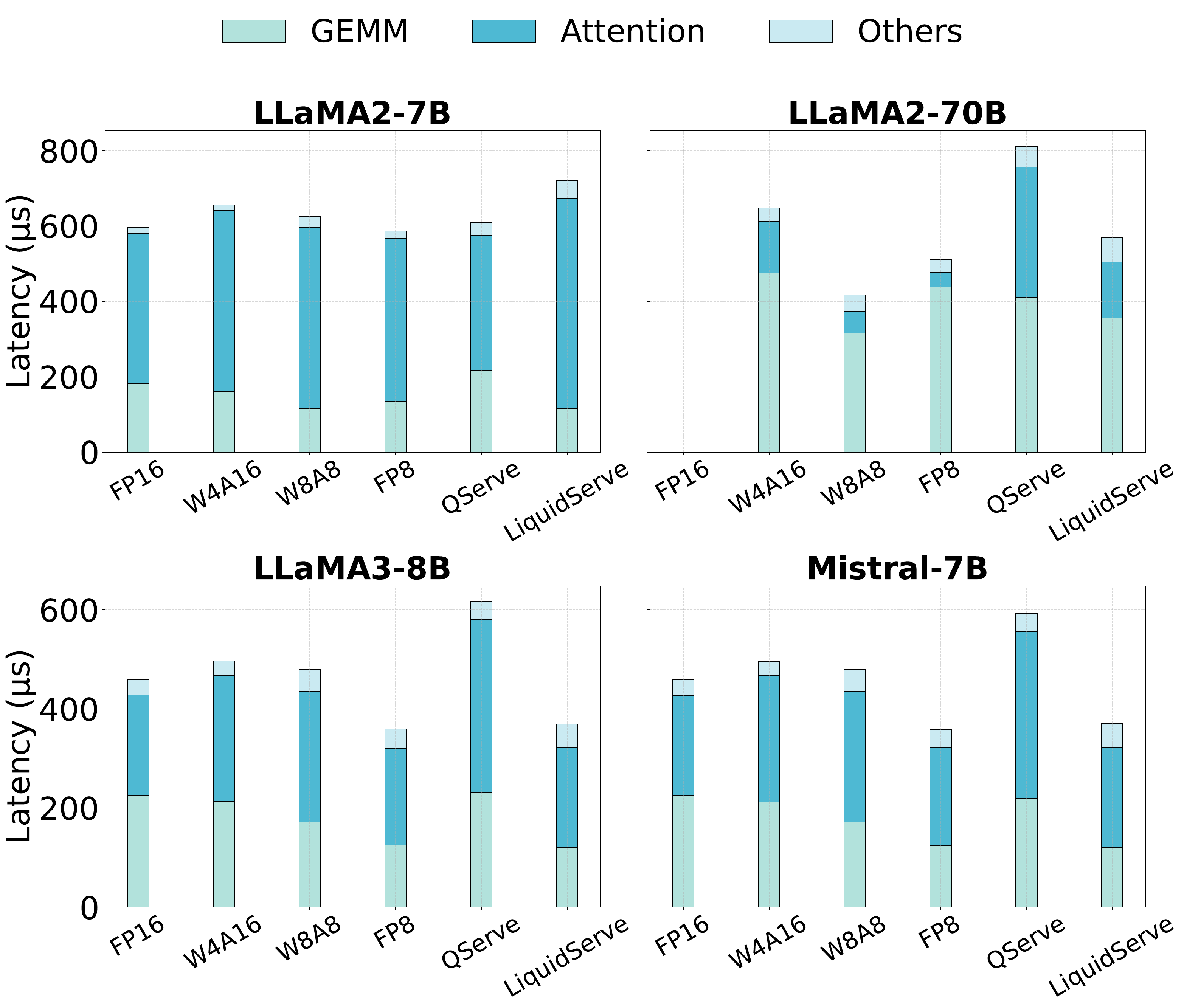}
    \centering
    \caption{Time breakdown for processing one decoding layer of LLaMA2-7B, LLaMA2-70B, LLaMA3-8B, and Mistral-7B at the batch sizes specified in Table~\ref{tab:max_throughput_result}.}
    \label{fig:end2end_time_breakdown}
\end{figure}

\vspace{3pt}
\noindent\textbf{Time Breakdown of End-to-End LLM Serving.} To analyze the sources of end-to-end performance gains in Table~\ref{tab:max_throughput_result}, we break down the inference time of one layer into GEMM, Attention, and Others for LLaMA2-7B, LLaMA2-70B, LLaMA3-8B, and Mistral-7B at the corresponding batch sizes. Note that while larger batch sizes generally yield higher throughput, they also increase the per-layer processing workload. As shown in Figure~\ref{fig:end2end_time_breakdown}, LiquidServe consistently delivers GEMM latencies that are on par with or better than all baselines, thanks to LiquidGEMM’s hardware-friendly dequantization that mitigates the CUDA kernel bottleneck.

On LLaMA2-7B, LiquidServe achieves the lowest GEMM latency, 1.90x faster than QServe and up to 1.58x faster than TRT. On LLaMA2-70B, despite using larger batch sizes, it still outperforms QServe by 1.15x, though slightly slower than TRT-W8A8. For LLaMA3-8B and Mistral-7B, LiquidServe matches FP8 in GEMM latency but incurs slightly higher overhead in the “Others” category. These results highlight LiquidGEMM’s effectiveness in improving overall inference efficiency.

\begin{figure}[t]\small
    \setlength{\abovecaptionskip}{3pt}
    \setlength{\belowcaptionskip}{0pt}
    \includegraphics[width=1.0\linewidth]{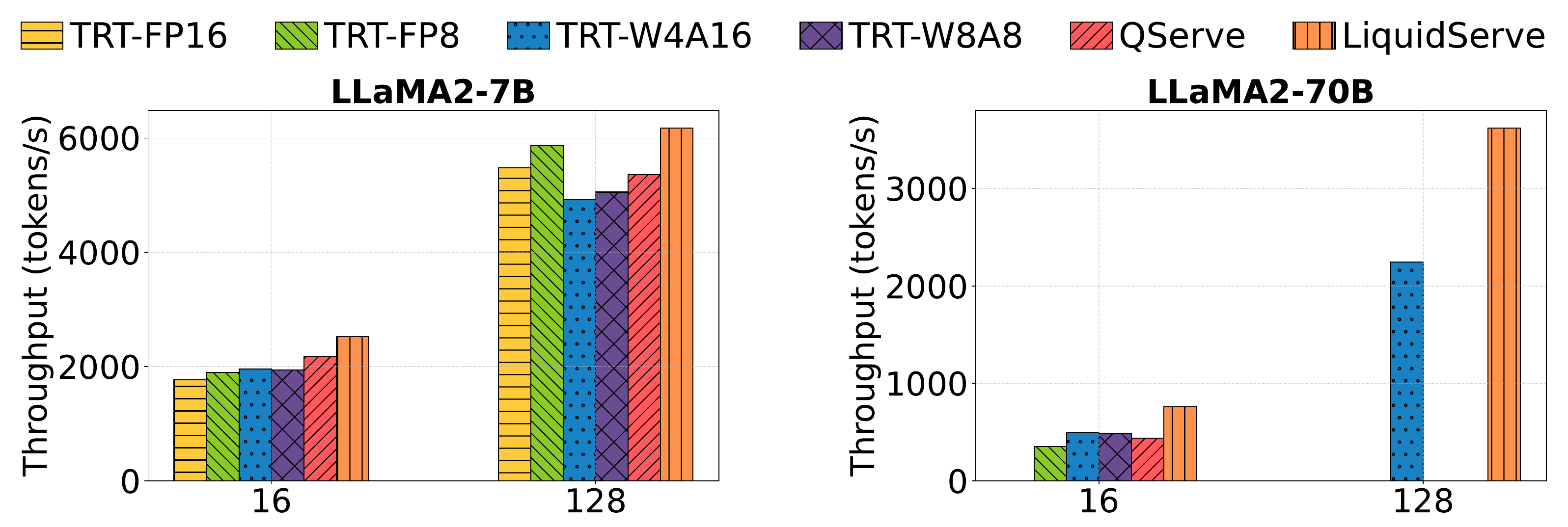}
    \centering
    \caption{Comparison of token generation throughput across systems at the same batch size.}
    \label{fig:fix_bs_throughput}
\end{figure}

\vspace{3pt}
\noindent\textbf{Throughput at Fixed Batch Sizes.} To further evaluate system performance under consistent workload settings, we compare the throughput of different systems using the same batch size, with LLaMA2-7B and LLaMA2-70B as representatives due to space constraints. Figure~\ref{fig:fix_bs_throughput} shows the token throughput for two batch sizes: 16, which is generally memory-bound, and 128, which approaches compute-bound. Missing bars indicate out-of-memory failures. LiquidServe consistently outperforms all baselines, demonstrating its efficiency advantage.

\subsection{Efficiency Comparison of GEMM Kernel}

We evaluate the efficiency of quantized GEMM kernels from all systems using a unified test framework. Specifically, we measure the compute performance on all GEMMs of a single-layer transformer across various models, including the fused QKV projection GEMM, the output projection GEMM, and the two FFN GEMMs, averaging the results over five runs.

\begin{figure}[t]
	\setlength{\abovecaptionskip}{0pt}
	\setlength{\belowcaptionskip}{0pt}
		\captionsetup[subfigure]{aboveskip=0pt,belowskip=0pt}
	\centering

    \includegraphics[width=0.9\columnwidth]{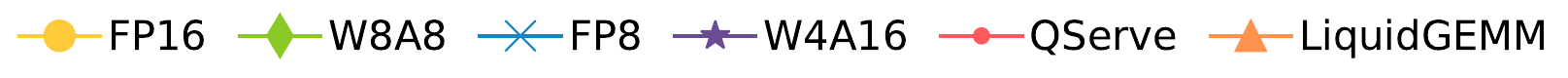}
	\begin{subfigure}[t]{0.23\textwidth}
            \captionsetup{position=top}
		\centering
		\includegraphics[width=\textwidth]{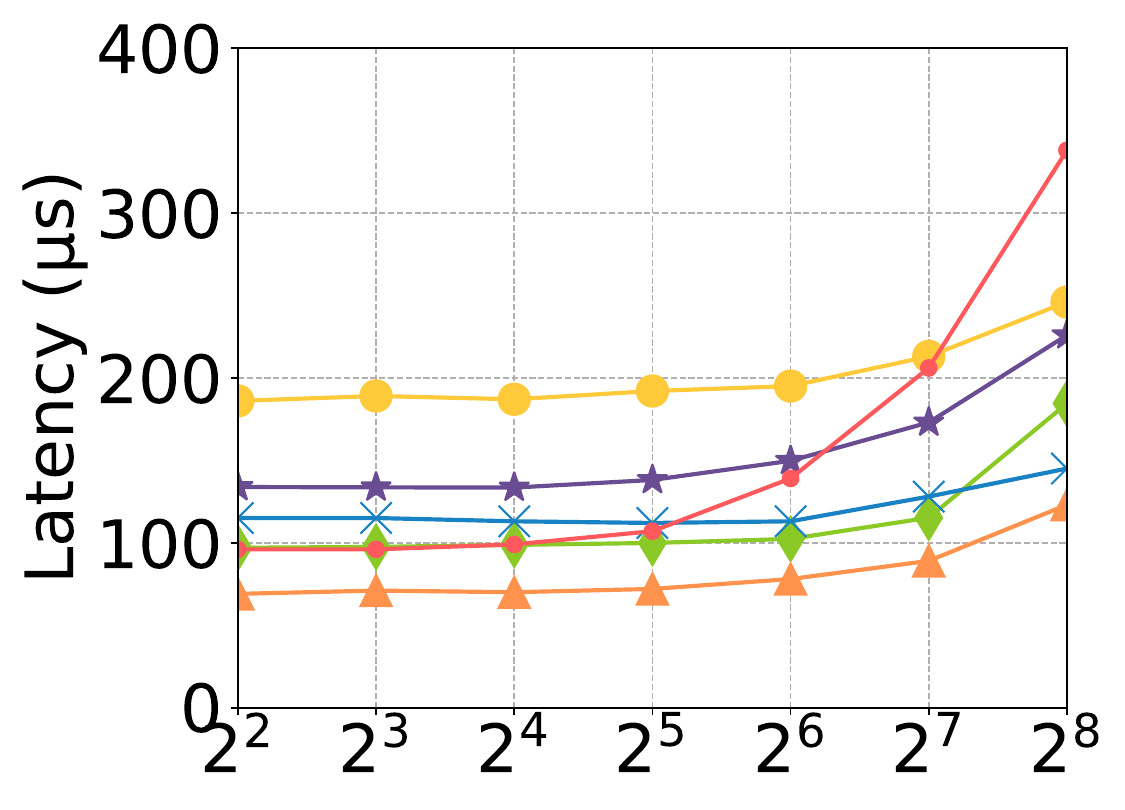}
		\caption{LLaMA2-7B}
		\label{fig:evaluation_gemm_latency_llama2-7b}
	\end{subfigure}
        \begin{subfigure}[t]{0.23\textwidth}
                \captionsetup{position=top}
    		\centering
    		\includegraphics[width=\textwidth]{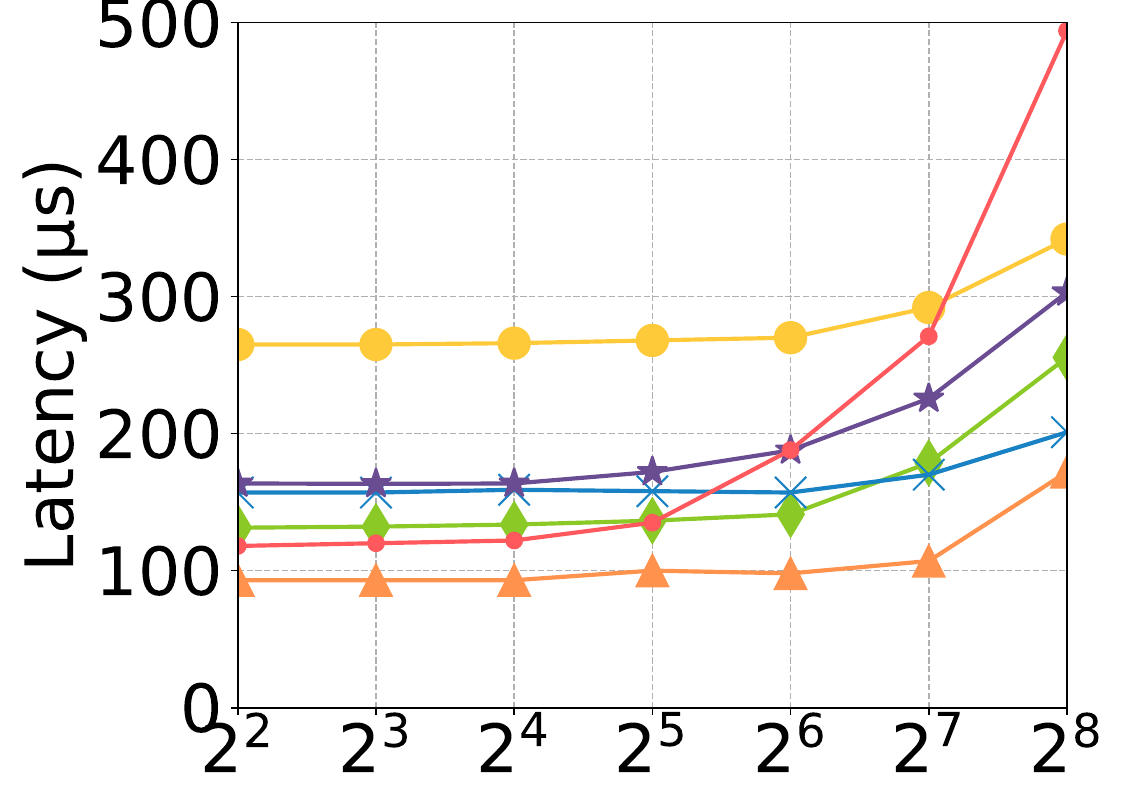}
    		\caption{LLaMA2-13B}
    		\label{fig:evaluation_gemm_latency_llama2-13b}
    	\end{subfigure}
        \begin{subfigure}[t]{0.23\textwidth}
                \captionsetup{position=top}
    		\centering
    		\includegraphics[width=\textwidth]{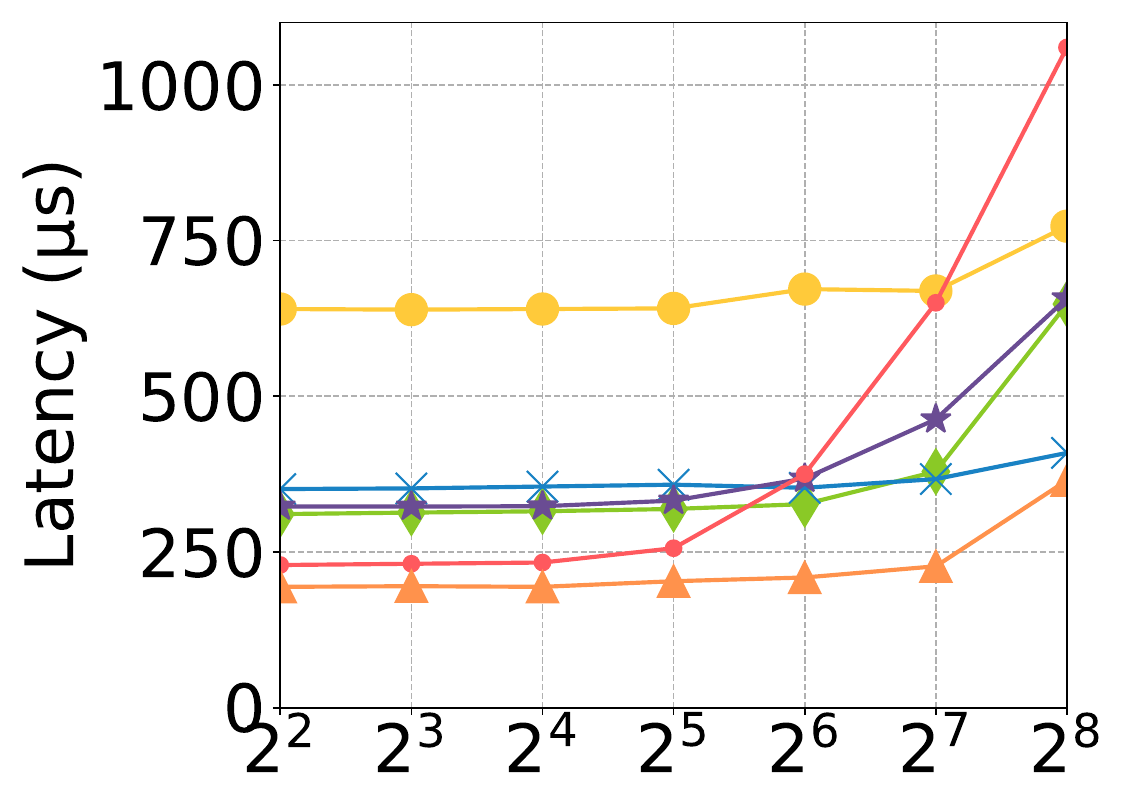}
    		\caption{LLaMA2-70B}
    		\label{fig:evaluation_gemm_latency_llama2-70b}
    	\end{subfigure}
        \begin{subfigure}[t]{0.23\textwidth}
                \captionsetup{position=top}
    		\centering
    		\includegraphics[width=\textwidth]{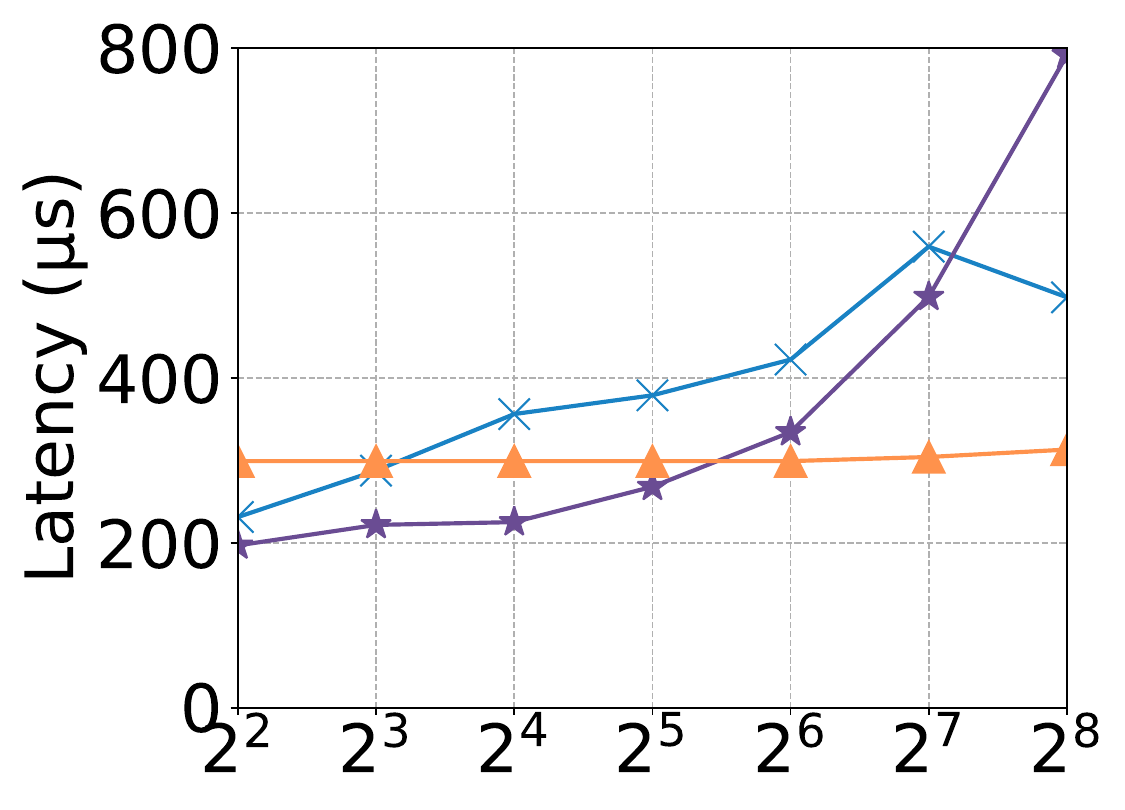}
    		\caption{Mixtral-8$\times$7B}
    		\label{fig:evaluation_gemm_latency_mixtral-8x7b}
    	\end{subfigure}
	\caption{Comparison of GEMM latency on the FFN layer with batch sizes ranging from 4 to 256.}
	\label{fig:gemm_latency}
\end{figure}

\vspace{3pt}
\noindent\textbf{Efficiency Comparison.}
Figure~\ref{fig:gemm_latency} compares GEMM latency across batch sizes from 4 to 256. On LLaMA models, QServe and LiquidGEMM generally outperform other systems on LLaMA2-13B and LLaMA2-70B at small batch sizes due to the benefits of 4-bit quantization. However, as batch size increases, QServe experiences significant performance degradation, while LiquidGEMM maintains consistently lower latency. At batch size 256, LiquidGEMM achieves speedups of 2.75x, 2.87x, and 2.90x over QServe on LLaMA2-7B, 13B, and 70B, respectively. For Mixtral-8×7B, TRT-W4A16 and TRT-FP8 outperform LiquidGEMM when the batch size is below 32, as they utilize a specialized GEMV kernel optimized for small-batch scenarios. Beyond batch size 32, LiquidGEMM delivers 1.41–1.84x speedup over TRT-FP8 and 1.12–2.53x over TRT-W4A16, demonstrating strong scalability and robustness.

\begin{figure}[t]\small
    \setlength{\abovecaptionskip}{3pt}
    \setlength{\belowcaptionskip}{0pt}
    \includegraphics[width=1.0\linewidth]{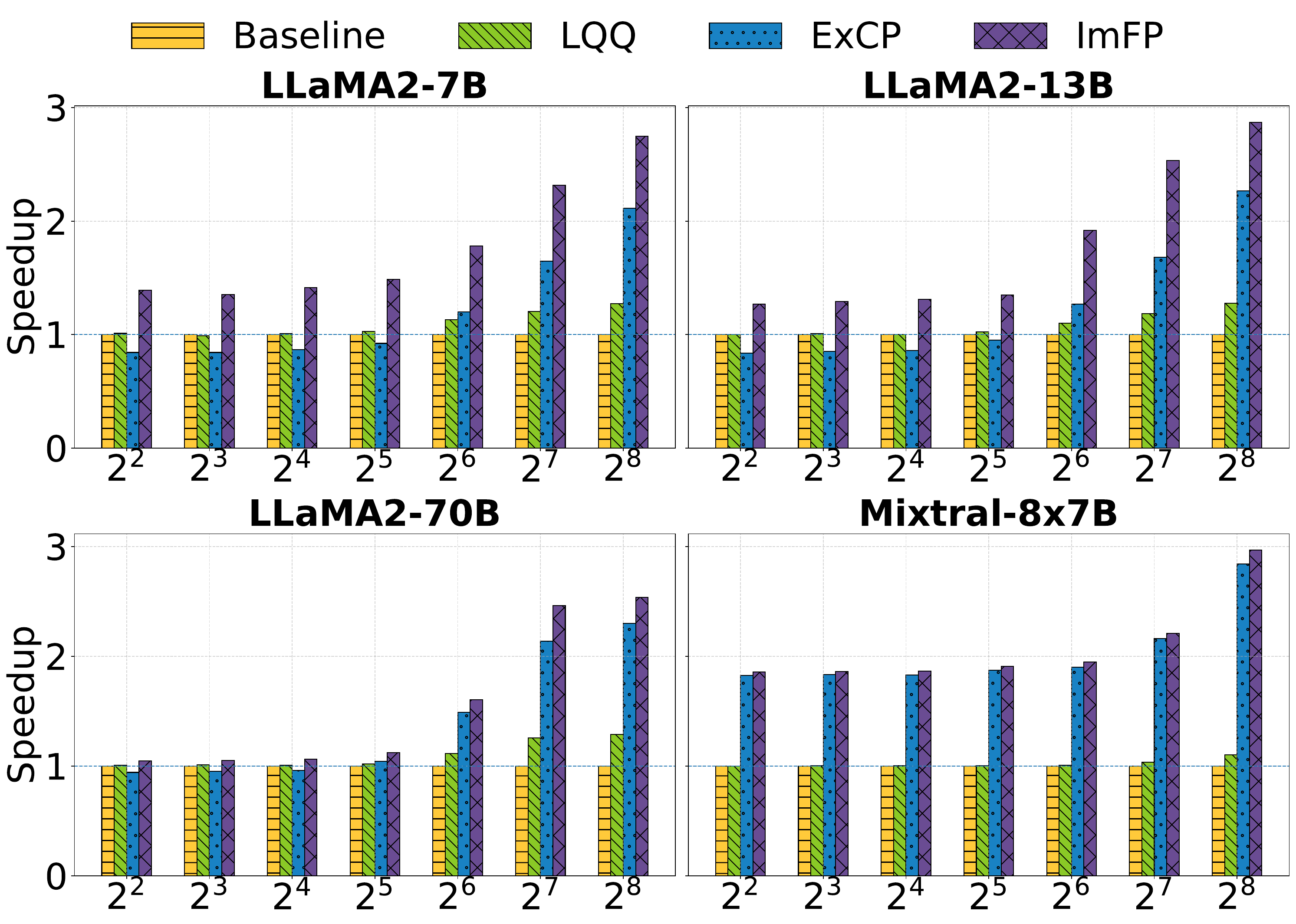}
    \centering
    \caption{Ablation study of LiquidGEMM by first enabling LQQ, followed by either the explicit coarse-grained pipeline (ExCP) or the implicit fine-grained pipeline (ImFP).}
    \label{fig:gemm_latency_ablation}
\end{figure}

\vspace{3pt}
\noindent\textbf{Ablation Study.} Figure~\ref{fig:gemm_latency_ablation} presents the ablation results. We first enable LQQ. When the batch size is small, GEMM is memory-bound and LQQ offers limited benefit. As batch size increases and computation becomes dominant, LQQ yields up to 1.29x speedup. Enabling ExCP at small batch sizes degrades performance due to round-trip traffic and synchronization overhead. At larger batch sizes, pipeline execution becomes more effective, and ExCP begins to provide benefits. In contrast, ImFP consistently improves performance across all batch sizes. Note that both ExCP and ImFP share the same memory layout and dequantization logic. Their advantage arises from pipelined execution across grouped GEMMs, particularly for MoE models, while the baseline and LQQ-only variants lack such inter-GEMM pipelining. In summary, these results highlight the effectiveness of the LQQ algorithm and the superior performance of the ImFP pipeline strategy.

\section{Related Work} \label{sec:related_work}

\noindent\textbf{Quantization for LLM Inference.} The methods typically fall into two categories: the weight-only quantization and the weight-activation quantization. For weight-only quantization, GPTQ~\cite{frantar2022gptq} pioneered sub-8-bit quantization by compressing weights to 3 or 4 bits using approximate second-order information. AWQ~\cite{lin2024awq} further improved accuracy by incorporating activation statistics to identify and preserve critical weights. For weight-activation quantization, GPT3.int8()~\cite{dettmers2022gpt3} introduced mixed-precision quantization to isolate outlier activations in separate 16-bit multiplications. SmoothQuant~\cite{xiao2023smoothquant} proposed migrating quantization challenges from activations to weights via mathematically equivalent transformations, effectively smoothing activation outliers. Atom~\cite{zhao2024atom} employed mixed-precision fine-grained group quantization, achieving a balance between throughput and accuracy. OmniQuant~\cite{shao2023omniquant} offered methods to automatically learn optimal quantization parameters. QServe~\cite{lin2024qserve} utilized a two-stage W4A8KV4 quantization approach optimized for GPU Tensor Cores, while QQQ~\cite{zhang2024qqq} combined adaptive smoothing with Hessian-based compensation, developing a new W4A8 GEMM kernel. Other recent works, including QuaRot~\cite{ashkboos2024quarot}, SpinQuant~\cite{liu2024spinquant}, and DuQuant~\cite{lin2025duquant}, applied transformations (e.g., rotations) to distribute outliers effectively. Among these, DuQuant simplified training complexity compared to SpinQuant and demonstrated superior performance over QuaRot. Different from these works, this paper focuses on the efficiency of W4A8 GEMM for efficient LLM serving.

\vspace{3pt}
\noindent\textbf{Quantization for LLM Training.} Quantization-aware training (QAT) achieves higher accuracy than post-training quantization (PTQ), but it has seen limited use due to its high computational overhead. LLM-QAT~\cite{liu2023llm} introduced a data-free distillation method using generations from pre-trained models, enabling practical QAT for LLMs. EfficientQAT~\cite{chen2024efficientqat} accelerated QAT with a two-phase training strategy. Bondarenko et al.~\cite{bondarenko2024low} proposed a lightweight, memory-efficient low-rank QAT method specifically for LLMs. EdgeQAT~\cite{shen2024edgeqat} introduced an entropy-guided approach with adaptive token importance to reduce information distortion in QAT.

\vspace{3pt}
\noindent\textbf{LLM Serving.} Orca~\cite{yu2022orca} optimized serving performance with iteration-level scheduling and selective batching. vLLM~\cite{kwon2023efficient} improved KV cache management efficiency through PagedAttention, inspired by virtual memory mechanisms. NVIDIA’s TensorRT-LLM~\cite{tensorrt_llm} provided an open-source library specifically optimized for accelerating LLM inference on GPUs. DistServe~\cite{zhong2024distserve} enhanced serving performance by disaggregating prefill and decoding computations with advanced placement algorithms. COMET~\cite{liu2024comet} proposed a mixed-precision inference framework incorporating novel, highly optimized kernels to maximize LLM inference performance.
\section{Conclusion} \label{sec:conclusion}

This paper addresses the dequantization bottleneck in W4A8 quantization for LLM serving. We first profile existing W4A8 GEMM kernels and develop a cost model to identify the critical performance factors. Guided by this analysis, we propose LiquidGEMM, a hardware-efficient W4A8 GEMM kernel that integrates two co-designed techniques: LiquidQuant, an overflow-safe dequantization algorithm; and an implicit fine-grained pipeline that maximizes parallelism across GPU subsystems. Experiments show up to 2.90x kernel speedup and 4.94x system-level speedup over prior W4A8 kernels, and 1.12–1.63x improvements over NVIDIA TensorRT-LLM. These results demonstrate that hardware-aware design makes W4A8 GEMM both efficient and scalable for high-performance LLM inference.

\bibliographystyle{ACM-Reference-Format}
\bibliography{references}
\end{document}